\documentclass[10pt,a4paper]{article}

\usepackage[english]{babel}
\usepackage[utf8]{inputenc} 
\usepackage[inner=3cm,outer=3cm,bottom=3cm,top=3cm]{geometry}
\usepackage{graphicx}
\usepackage{array}
\usepackage{morefloats}
\usepackage{amsfonts,amsmath,color}
\usepackage{xcolor}												
\usepackage{tikz}
\usepackage[T1]{fontenc}    
\usepackage{appendix}
\usepackage{stackrel}
\usepackage{eurosym}
\usepackage{subfigure}
\usepackage{epstopdf} 
\usepackage{amscd,amssymb,amsthm,latexsym, amsbsy,amsmath}
\usepackage{multicol}
\usepackage{multirow}
\usepackage{booktabs}
\usepackage{float}
\usepackage{color, colortbl}
\usepackage{xcolor}
\usepackage{enumerate} 
\usepackage[round]{natbib} 
\usepackage{rotating}  
\usepackage{textcomp} 
\usepackage{longtable} 
\usepackage{authblk} 

\definecolor{bluegray}{rgb}{0.4, 0.6, 0.8} 
\definecolor{darkpastelgreen}{rgb}{0.01, 0.75, 0.24} 
\definecolor{byzantine}{rgb}{0.74, 0.2, 0.64} 
\definecolor{darktangerine}{rgb}{1.0, 0.66, 0.07} 

\newcolumntype{T}[1]{>{\centering\let\newline\\\arraybackslash\hspace{0pt}}m{#1}}
\LTcapwidth=\textwidth

\providecommand{\keywords}[1]{\textbf{\textit{Key words---}} #1}

\title{A critical review of LASSO and its derivatives for variable selection under dependence among covariates}
\vspace{0.1cm}
\author[1]{Laura Freijeiro-González}
\author[1]{Manuel Febrero-Bande}
\author[1]{Wenceslao González-Manteiga}
\affil[1]{Department of Statistics, Mathematical Analysis and Optimization; Santiago de Compostela University, Santiago de Compostela, Spain. Email: laura.freijeiro.gonzalez@usc.es}
\date{}                     
\begin{document}
	

\maketitle

\begin{abstract}
	We study the limitations of the well known LASSO regression as a variable selector when there exists dependence structures among covariates. We analyze both the classic situation with $n\geq p$ and the high dimensional framework with $p>n$. Restrictive properties of this methodology to guarantee optimality, as well as the inconveniences in practice, are analyzed. Examples of these drawbacks are showed by means of a extensive simulation study, making use of different dependence scenarios. In order to search for improvements, a broad comparison with LASSO derivatives and alternatives is carried out. Eventually, we give some guidance about what procedures are the best in terms of the data nature. 
\end{abstract}	



\section{Introduction and motivation}

Nowadays, in many important statistical applications, it is of high relevance to apply a first variable selection step to correctly explain the data and avoid unnecessary noise. Furthermore, it is usual to find that the number of variables $p$ is larger than the number of available samples $n$ ($p>n$). Some examples of fields where this framework arises are processing image, statistical signal processing, genomics or functional magnetic resonance imaging (fMRI) among others. It is in the $p>n$ context where the ordinary models fail and, as a result, estimation and prediction in these settings are generally acknowledged as an important challenge in contemporary statistics.

In this framework, one of the most studied fields is the regression models adjustment. The idea of a regression model is to explain a variable of interest, $Y$, using $p$ covariates $X_1,\dots,X_p$. This is done by means of a structure $m(X)$, $X=(X_1,\dots,X_p)^\top\in \mathbb{R}^p$, and an unknown error $\varepsilon$:
\begin{equation}\label{regression_model}
Y=m(X)+\varepsilon.
\end{equation}
Here, $m(X)$ denotes the type of relation between the dependent variable $Y$ and the $p$ explanatory covariates, while $\varepsilon$ is the term which captures the remaining information as well as other unobserved fluctuations. This is typically assumed to have null mean and variance $\sigma^2$.

Once the $m(X)$ structure of (\ref{regression_model}) is estimated, it is possible to know the importance of every $X_1,\dots,X_p$ in terms of explaining $Y$ apart from making predictions. Nevertheless, this estimation when $p>n$ still is a difficult and open problem in many situations. In particular, its easiest expression: the linear regression model, has been widely studied in the last years so as to provide efficient algorithms to fit this (see for example \cite{giraud2014introduction} or \cite{hastie2015statistical}).

In linear regression, as the name suggests, the relationship between $Y$ and $X$ is assumed to be linear, giving place to the model:
\begin{equation}\label{linear_regre}
Y=X\beta+\varepsilon,
\end{equation}
where $\beta\in\mathbb{R}^p$ is a coefficients vector to estimate. Note that we assume the covariates $X_1,\dots,X_p$ and the response $Y$ centered, excluding the intercept from the model without loss of generality.

Having data $(y_i,\mathbf{x_i})\in \mathbb{R}^{p+1}$ for $i=1,\dots,n$ samples, denoting $y=(y_1,\dots,y_n)^\top\in \mathbb{R}^n$ and $\mathbf{X}=(\mathbf{x_1},\dots,\mathbf{x_p})\in\mathbb{R}^{n\times p}$ with $\mathbf{x_j}=(x_{ij})_{i=1}^n \in \mathbb{R}^{n}$ for $j=1,\dots,p$, the $\beta$ vector can be estimated using the classical ordinary least squares (OLS) method solving (\ref{OLS_problem}). 
\begin{equation}\label{OLS_problem}
\hat{\beta}^{OLS}=\min_{\beta} \left\lbrace \sum_{i=1}^{n}\left(y_i-\sum_{j=1}^{p}x_{ij}\beta_j\right)^2 \right\rbrace 
\end{equation}
This estimator, $\hat{\beta}^{OLS}=(\mathbf{X}^\top\mathbf{X})^{-1}\mathbf{X}^\top y$, enjoys some desirable properties such as being an unbiased and consistent $\beta$ estimator of minimum variance.

Nevertheless, when $p>n$, this estimation method fails, as there are infinite solutions for the problem (\ref{OLS_problem}). Then, it is necessary to impose modifications on the procedure or to consider new estimation algorithms able to recover the $\beta$ values.

In order to overcome this drawback, the LASSO regression (\cite{Tibshirani1996}) is still widely used due to its capability of reducing the dimension of the problem. This methodology assumes sparsity in the coefficient vector $\beta$, resulting in an easier interpretation of the model and performing variable selection. However, some rigid assumptions on the covariates matrix and sample size are needed so as to guarantee its good behavior (see, for example, \cite{Meinshausen2010}). Moreover, the LASSO procedure exhibits some drawbacks related to the correct selection of covariates and the exclusion of redundant information (see \cite{Su2017}), aside from bias. This can be easily showed in controlled simulated scenarios where it is known that only a small part of the covariates are relevant.

Hence, is this always the best option or at least a good start point? We have not found a totally convincing answer to this question in the literature. So as to test its performance and extract some general conclusions, the characteristics of the LASSO procedure are analyzed in this article. For this purpose, we start revisiting the existing literature about this topic as well as its most important adaptations. Furthermore, in view of the LASSO limitations, a global comparison is developed so as to test which procedures are capable of overcoming these in different dependence contexts, comparing their performance with alternatives which have proved their efficiency. Finally, some broad conclusions are drawn.

The article is organized as follows, in Section \ref{sec:LASSO} a complete overview of the LASSO regression is given, including a summary of the requirements and inconveniences this algorithm has to deal with. In Section \ref{simulation_scenarios}, some special simulation scenarios are introduced and used to illustrate the problems of this methodology in practice, testing the behavior of the LASSO under different dependence structures. In Section \ref{other} the evolution of the LASSO in the last years is analyzed. Besides, other efficient alternatives in covariates selection are briefly described and their performance is compared with the LASSO one. Eventually, in Section \ref{conclusions},
a discussion is carried out so as to give some guidance about what types of covariates selection procedures are the best ones in terms of the data dependence structure.

\section{A complete overview of the LASSO regression}\label{sec:LASSO}

In a linear regression model as the one of (\ref{linear_regre}), there are a lot of situations where not all $p$ explanatory covariates are relevant, but several are unnecessary. In these scenarios we can assume that the $\beta$ vector is sparse and then search for the important covariates, avoiding noisy ones. The idea is, somehow, to obtain a methodology able to compare the covariates and select only those most important, discarding irrelevant information and keeping the error of prediction as small as possible. As there are $2^p$ possible sub-models, it tends to be rather costly to compare all of them using techniques such as forward selection or backward elimination.

One of the most typical solutions is to impose a restriction on the number of included covariates. This is done by means of adding some constraints to the OLS problem (\ref{OLS_problem}).

This brings up the idea of a model selection criterion, which express a trade-off between the goodness of fit and the complexity of the model, such as the AIC (\cite{akaike1998information}) or BIC (\cite{schwarz1978estimating}). Nevertheless, these approaches are computationally intensive, hard to derive sampling properties and unstable. As a result, they are not suitable for scenarios where the dimension of $p$ is large.

Therefore, we could think in penalizing the irrelevant information by means of the number of coefficients included in the final model. This can be done adding a penalty factor $p_\lambda (\beta)$ in (\ref{OLS_problem}), resulting in the problem
\begin{equation}\label{penalization_problem}
\min_{\beta} \left\lbrace \sum_{i=1}^{n}\left(y_i-\sum_{j=1}^{p}x_{ij}\beta_j\right)^2+ p_\lambda (\beta ) \right\rbrace .
\end{equation} 

For this purpose, following the ideas of goodness-of-fit measures, a $L_0$ regularization, $\lambda \|\beta\|_0=\lambda \sum_{j=1}^{p}  \mathbf{1}_{\beta_j \not=0}$, could be applied. This criterion penalizes models which include more covariates but do not improve too much the performance. This results in a model with the best trade-off between interpretability and accuracy, as the AIC or BIC criterion philosophy does, obtaining
\begin{equation}
\label{bL0}
\hat{\beta}^{L_0}=  \min_{\beta} \left\lbrace \sum_{i=1}^{n}\left(y_i-\sum_{j=1}^{p}x_{ij}\beta_j\right)^2+ \lambda\sum_{j=1}^{p} \mathbf{1}_{\beta_j \not=0} \right\rbrace ,
\end{equation}
where $\lambda>0$ is a regularization parameter.

The problem (\ref{bL0}) is known as the best subset selection (\cite{beale1967discarding}, \cite{hocking1967selection}). This is non-smooth and non-convex, which hinders to achieve an optimal solution. Then, the estimator $\hat{\beta}^{L_0}$ is infeasible to compute when $p$ is of medium or large size, as (\ref{bL0}) becomes a $NP$-hard problem with exponential complexity. See \cite{hastie2017extended} for a comparison of this procedure with more current methods.

\begin{figure}[htb]\centering
	\includegraphics[width=0.6\linewidth]{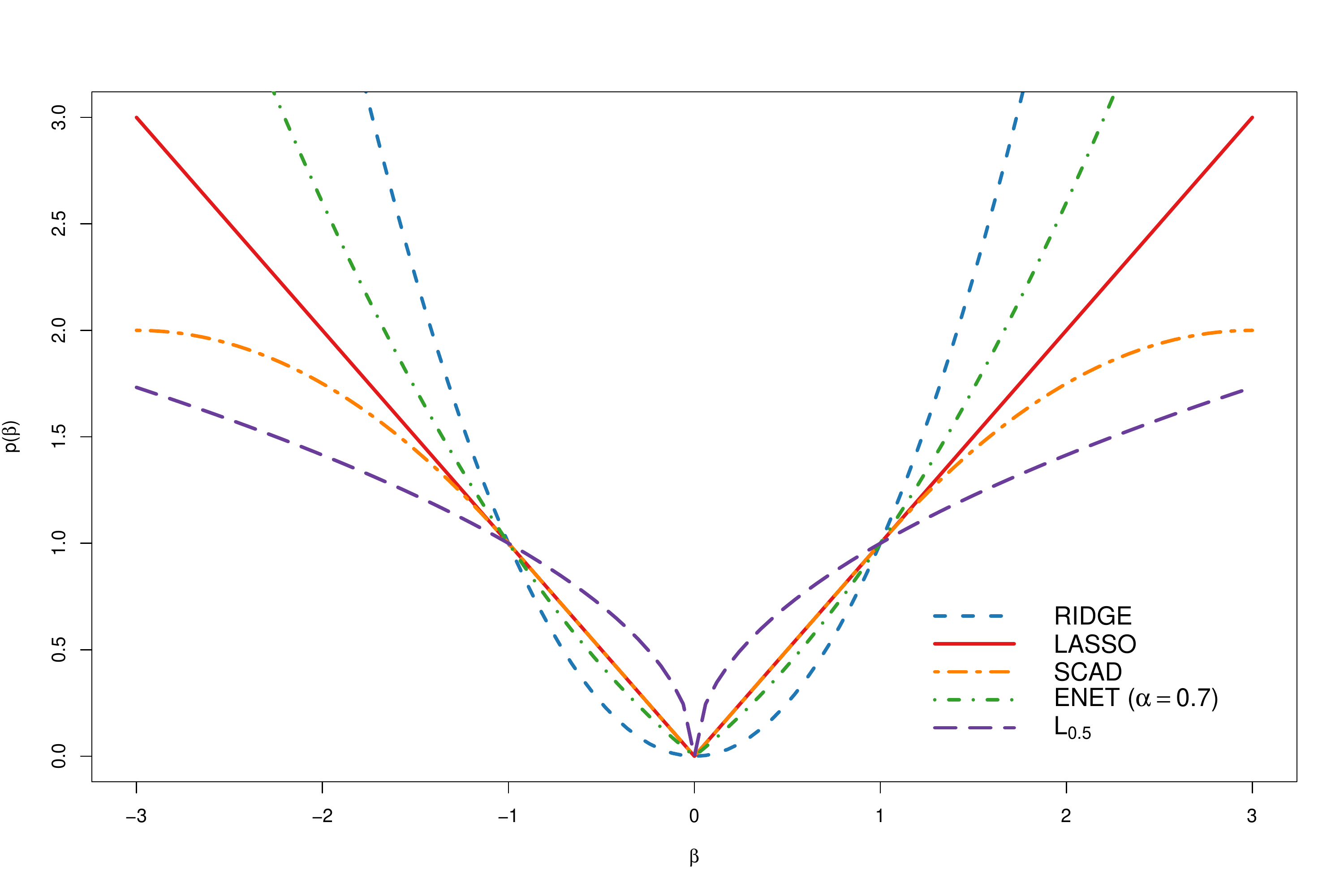}
	\caption[Comparison of different penalization methods: $L_2$ or RIDGE penalization ($RIDGE$), $L_1$ or LASSO penalization ($LASSO$), SCAD regularization ($SCAD$), Elastic Net penalization method for $\alpha=0.7$ ($ENET  (\alpha=0.7)$) and $L_{0.5}$ regularization ($L_{0.5}$)]{\label{penalizations_draw} Comparison of different penalization methods: $L_2$ or RIDGE penalization ($RIDGE$), $L_1$ or LASSO penalization ($LASSO$), SCAD regularization ($SCAD$), Elastic Net penalization method for $\alpha=0.7$ ($ENET \; (\alpha=0.7)$) and $L_{0.5}$ regularization ($L_{0.5}$) \footnotemark[1].}
\end{figure}

So, to avoid this drawback, it is possible to replace $\lambda \|\beta\|_0$ by other types of penalization. Taking into account that this belongs to the family $p_\lambda (\beta_j )=\lambda \|\beta\|_q := \lambda \left( \sum_{j=1}^{p} \sqrt[q]{|\beta_j|} \right)^q$, with $q \geq 0$, we can commute this for a more appropriate one. The problem (\ref{penalization_problem}) with this type of penalization is known as the bridge regression (\cite{fu1998penalized}). The caveat of this family is that this only selects covariates for the values $1\geq q >0$. Moreover, the problem (\ref{penalization_problem}) is only convex for the $q=1$ case (see Figure \ref{penalizations_draw}). Then, it seems reasonable to work with the norm $\|\beta\|_1=\sum_{j=1}^{p}|\beta_j|$, which is convex, allows covariates selection and leads to the extensively studied LASSO (Least Absolute Shrinkage and Selection Operator) regression, see \cite{Tibshirani1996} and \cite{tibshirani2011regression}.

\footnotetext[1]{ Some of these procedures will be introduced later in Section \ref{other} }

The LASSO, also known as basis pursuit in image processing (\cite{chen2001atomic}, \cite{candes2006stable}, \cite{donoho2005stable}), was presented by \cite{Tibshirani1996}. This proposes the imposition of a $L_1$ penalization in (\ref{OLS_problem}) with the aim of performing variable selection and overcoming the high dimensional estimation of $\beta$ drawback when $p>n$. In this way, it would be needed to solve the optimization problem
\begin{equation*} 
\begin{split}
&\hat{\beta}^{L_1}= \min_{\beta}  \sum_{i=1}^{n}\left(y_i-\sum_{j=1}^{p}x_{ij}\beta_j\right)^2,\\
&\text{subject to}  \sum_{j=1}^{p}|{\beta_j}|\leq t,
\end{split}
\end{equation*}
which can be rewritten like
\begin{equation}
\label{bLASSO}
\hat{\beta}^{L_1}=  \min_{\beta} \left\lbrace \sum_{i=1}^{n}\left(y_i-\sum_{j=1}^{p}x_{ij}\beta_j\right)^2+ \lambda\sum_{j=1}^{p}|{\beta_j}| \right\rbrace .
\end{equation}

The problem (\ref{bLASSO}) is convex, which guarantees that has always one solution at least, although if $p>n$, there may be multiple minimums (see \cite{tibshirani2013LASSO} for more details). Besides, assuming the noise term $\varepsilon$ to be Gaussian, $\hat{\beta}^{L_1}$ can be interpreted as a penalized maximum likelihood estimate in which the fitted coefficients are penalized in a $L_1$ sense. As a result, this encourages sparsity.

In these problems, the term $\lambda>0$, or $t>0$ equivalently, is the shrinkage parameter. For large values of $\lambda$ (small values of $t$) the coefficients of $\beta$ are more penalized, which results in a bigger number of elements shrinkaged to zero. Nevertheless, the estimator $\hat{\beta}^{L_1}$ of (\ref{bLASSO}) has not got an explicit expression.

The LASSO defined in (\ref{bLASSO}) can be viewed as a convex relaxation of the optimization problem with the $L_0$ analogue of a norm in (\ref{bL0}). Then, the requirement of computational feasibility and statistical accuracy can be met by this estimator.

This method has been widely studied over the last years: it has been showed that this procedure is consistent in terms of prediction (see \cite{van2009conditions} for an extensive analysis) and this guarantees consistency of the parameter estimates at least in a $L_2$ sense (\cite{van2009conditions}, \cite{meinshausen2009LASSO}, \cite{candes2007dantzig}); besides, this is a consistent variable selector under some assumptions (\cite{Meinshausen2006}, \cite{wainwright2009sharp}, \cite{Zhao2006}).

\subsection{Analysis of the LASSO regression requirements and inconveniences}\label{problems_LASSO}

In spite of all these good qualities, the LASSO regression has some important limitations in practice (see for example \cite{Zou2005} or \cite{Su2017}). These limitations are analyzed in the next subsections, collecting some recent developed theoretical properties and displaying how far it is possible to ensure its good behavior.

\subsubsection{Biased estimator}\label{biased_estimator}

In the context of having more covariates $p$, than number of samples $n$, the LASSO regression can identify at most $n$ important covariates before it saturates (see \cite{Zou2005}). This restriction is common for almost all regression adjustment methods which rely on penalizations in this framework. Specially for those based on $L_1$ ideas. As a result, an estimator $\hat{\beta}$ can have at most $n$ coefficients not equal zero. Into words, there is not enough available information to adjust the whole model. This situation can be compared with a system of equations in which we have more variables than equations per se.

Related with this, another caveat of penalization processes is the bias. This produces higher prediction errors. In the LASSO adjustment, the imposition of the $L_1$ penalization in the OLS problem (\ref{OLS_problem}) as a safe passage to estimate $\beta$ has a cost, which is translated in bias (see \cite{hastie2009elements}, \cite{giraud2014introduction} or \cite{hastie2015statistical}). This can be easily explained under orthogonal design, where the $L_1$ penalization results in a perturbation of the unbiased OLS estimator $\hat{\beta}^{OLS}$ given by
\begin{equation}\label{bLASSO_bols}
\hat{\beta}^{L_1}_j=\textrm{sign}(\hat{\beta}_j^{OLS})(|\hat{\beta}_j^{OLS}|-\lambda)_+,
\end{equation}
where $\textrm{sign}(\cdot)$ denotes the sign of the coefficients and $(\cdot)_+$ equals to zero all quantities which are not positive. This results in a soft threshold of the ordinary mean square estimator ruled by the $\lambda>0$ parameter, where the coefficients $|\hat{\beta}^{OLS}_j|\leq\lambda$ are adjusted to zero.

In order to correct the bias, weighted versions of the LASSO method based on iterative schemes, have been developed. An example is the popular adaptive LASSO (\cite{Zou2006}, \cite{huang2008adaptive}, \cite{Geer2011}). This procedure gives different weights to each covariate in the penalization part, readjusting these in every step of the iterative process until convergence.

\subsubsection{Consistency of the LASSO: neighborhood stability condition}\label{consistency}

Despite the LASSO is broadly employed, it is not always possible to guarantee its proper performance. As we can see in \cite{buhlmann2011statistics}, certain conditions are required to guarantee an efficient screening property for variable selection. However, this presents some important limitations as a variable selector when these do not hold.

For example, when the model has several highly correlated covariates with the response, LASSO tends to pick randomly only one or a few of them and shrinks the rest to $0$ (see \cite{Zou2005}). This fact results in a confusion phenomenon if there are high correlations between relevant and unimportant covariates, and in a loss of information when the subset of important covariates have a strong dependence structure. Some algorithms which result in non-sparse estimators try to relieve this effect, like the Ridge regression (\cite{hoerl1970ridge}) or the Elastic Net (\cite{Zou2005}). An interpretation of their penalties is displayed in Figure \ref{penalizations_draw}.

Denoting $S=\{j:\beta_j\not=0\}$ the set of non-zero real values and estimating this by $\hat{S}$, this last would be a consistent estimator if this verifies 
\begin{equation}\label{consistet_LASSO_P}
\mathbb{P}(\hat{S}=S) \stackbin[n\rightarrow \infty]{}{\longrightarrow} 1.
\end{equation}

The condition (\ref{consistet_LASSO_P}) places a restriction on the growth of the number $p$ of variables and sparsity $|S|$, typically of the form $|S|\log(p)=o(n)$ (see \cite{Meinshausen2006}). Denoting $s=|S|$, this forces the necessity of $n>s\log(p)$ in order to achieve consistency for the LASSO estimator.

Besides, for consistent variable selection using $\hat{S}^{L_1}=\{j:\hat{\beta}^{L_1}_j\not=0\}$, it turns out that the design matrix of the model, $X$, needs to satisfy some assumptions. The strongest of which is arguably the so-called ``neighborhood stability condition'' (\cite{Meinshausen2006}). This condition is equivalent to the irrepresentable condition (\cite{Zhao2006}; \cite{Zou2006}; \cite{Yuan2007}):
\begin{equation}\label{irrepresent_cond}
\max_{j\in S^c} |\textrm{sign}(\beta_S)^\top (X_S^\top X_S)^{-1} X_S^\top X_j| \leq \theta \quad \text{for some } 0<\theta<1,
\end{equation}
being $\beta_S$ the subvector of $\beta$ and $X_S$ the submatrix of $X$ considering the elements of $S$.

If this condition is violated, all that we can hope for is recovery of the regression vector $\beta$ in an $L_2$-sense of convergence by achieving $\parallel\hat{\beta}^{L_1}-\beta \parallel_2 \stackbin[n\rightarrow \infty]{}{\longrightarrow_p} 0$ (see \cite{Meinshausen2010} for more details). Moreover, under some assumptions in the design, the irrepresentable condition can be expressed as the called ``necessary condition'' (\cite{Zou2006}). It is not an easy task to verify these conditions in practice, specially in contexts where $p$ can be huge.

Quoted \cite{buhlmann2011statistics}: roughly speaking, the neighborhood stability or irrepresentable condition (\ref{irrepresent_cond}) fails to hold if the design matrix $X$ is too much ``ill-posed'' and exhibits a too strong degree of linear dependence within ``smaller'' sub-matrices of X.

In addition, it is needed to assure that there are enough information and suitable characteristics for ``signal recovery'' of the sparse $\beta$ vector. This requires relevant covariates coefficients be large enough so as to distinguish them from the zero ones. Then, the non-zero regression coefficients need to satisfy 
\begin{equation}\label{beta_size}
\stackbin[j\in S]{}{\inf} |\beta_j| >> \sqrt{s \log(p)/n}	
\end{equation}
in order to guarantee the consistency of the $\hat{\beta}^{L_1}$ estimator of problem (\ref{bLASSO}). This is called a beta-min condition. Nevertheless, this requirement may be unrealistic in practice and small non-zero coefficients may not be detected (in a consistent way). See \cite{buhlmann2011statistics} for more information.

Owing to these difficulties, different new methodologies based on ideas derived from subsampling and bootstrap have been developed. Examples are the random LASSO (\cite{wang2011random}), an algorithm based on subsampling, or the stability selection method mixed with randomized LASSO of \cite{Meinshausen2010}. This last searches for consistency although the irrepresentable condition introduced in (\ref{irrepresent_cond}) would be violated.

\subsubsection{False discoveries of the LASSO}\label{false_discoveries}

As it is explained in \cite{Su2017}: In regression settings where explanatory variables have very low correlations and there are relatively few effects, each of large magnitude, we expect the LASSO to find the important variables with few errors, if any. Nevertheless, in a regime of linear sparsity, there exist a trade-off between false and true positive rates along the LASSO path, even when the design variables are stochastically independent. Besides, this phenomenon occurs no matter how strong the effect sizes are.

This can be translated as one of the major disadvantages of using LASSO like a variable selector is that exists a trade-off between the false discovery proportion (FDP) and the true positive proportion (TPP), which are defined as
\begin{equation}\label{FDP_TPP}
FDP(\lambda)=\frac{F(\lambda)}{|\{j: \hat{\beta}_j(\lambda)\not=0\}|\vee 1}\quad \text{and} \quad TPP(\lambda)=\frac{T(\lambda)}{s\vee 1},
\end{equation}
where $F(\lambda)=|\{j \in S^c: \hat{\beta}_j(\lambda)\not=0 \}|$ denotes the number of false discoveries, $T(\lambda)=|\{j \in S : \hat{\beta}_j(\lambda)\not=0\}|$ is the number of positive discoveries and $a\vee b=\max\{a,b\}$.

Then, it is unlikely to achieve high power and a low false positive rate simultaneously. Noticing that $FDP$ is a natural measure of type I error while $1-TPP$ is the fraction of missed signals (a natural notion of type II error), the results say that nowhere on the LASSO path can both types of error rates be simultaneously low. This also happens even when there is no noise in the model and the regressors are stochastically independent. Hence, there exists only a possible reason: it is because of the $L_1$ shrinkage which results in pseudo-noise. Furthermore, this does not occur with other types of penalizations, like the $L_0$ penalty.  See \cite{Su2017} for more details.

In fact, it can be proved in a quiet global context, that the LASSO is not capable of selecting the correct subset of important covariates without adding some noise to the model in the best case (see \cite{wasserman2009high} or \cite{Su2017}).

Then, modifications of the traditional LASSO procedure are needed in order to control the FDP. Some alternatives, such as the boLASSO procedure (see \cite{Bach2008}), which use bootstrap to calibrate the $FDP$, the thresholded LASSO (\cite{Zhou2010}), based on the use of a threshold to avoid noise covariates, or more recent ones, like the stability selection method (see \cite{Meinshausen2010}) or the use of knockoffs (see \cite{Hofner:StabSel:2015}, \cite{Weinstein2017}, \cite{candes2018panning} and \cite{barber2019knockoff}), were proposed to solve this drawback. To the best of our knowledge still there is not a version of this last for the $p>n$ framework.

\subsubsection{Correct selection of the penalization parameter $\boldsymbol{\lambda}$}\label{lambda_selection}
	
One of the most important parts of a LASSO adjustment is the proper selection of the penalization parameter $\lambda\geq0$. Its size controls both: the number of selected variables and the degree to which their estimated coefficients are shrunk to zero, controlling the bias as well. A too large value of $\lambda$ forces all coefficients of $\hat{\beta}^{L_1}$ to be null, while a value next to zero includes too many noisy covariates. Then, a good choice of $\lambda$ is needed in order to achieve a balance between simplicity and selection accuracy.

The problem of the proper choice of the $\lambda$ parameter depends on the unknown error variance $\sigma^2$. We can see in \cite{buhlmann2011statistics} that the oracle inequality states to select $\lambda$ of order $\sigma\sqrt{\log(p)/n}$ to keep the mean squared prediction error of LASSO as the same order as if we knew the active set $S$ in advance. In practice, the $\sigma$ value is unknown and its estimation with $p>n$ is quite complex. To give some guidance in this field we refer to \cite{fan2012variance} or \cite{reid2016study}, although this still is a growing study field.

Thus, other methods to estimate $\lambda$ are proposed. Following the classification of \cite{homrighausen2018study} we can distinguish three categories: minimization of a generalized information criteria (like AIC or BIC), by means of resampling procedures (such as cross-validation or bootstrap) or reformulating the LASSO optimization problem. Due to computational cost, the most used criteria to fit a LASSO adjustment are cross-validation techniques. Nevertheless, it can be showed that this criterion achieves an adequate $\lambda$ value for prediction risk but this leads to inconsistent model selection for sparse methods (see \cite{Meinshausen2006}). Then, for recovering the set $S$, a larger penalty parameter would be needed (\cite{buhlmann2011statistics}).

\cite{Su2017} argue that, when the regularization parameter $\lambda$ is needed to be large for a proper variable selection, the LASSO estimator is seriously biased downwards. The residuals still contain much of the effects associated with the selected variables, which is called shrinkage noise. As many strong variables get picked up, this gets inflated and its projection along the directions of some of the null variables may actually dwarf the signals coming from the strong regression coefficients, selecting null variables.

Nevertheless, to the best of our knowledge, there is not a common agreement about the way of choosing this $\lambda$ value. Hence, cross-validation techniques are widely used to adjust the LASSO regression. See \cite{homrighausen2018study} for more details.


\section{A comparative study with simulation scenarios}\label{simulation_scenarios}

Once the LASSO requirements have been introduced, its performance is tested in practice.  So, scenarios verifying and do not these conditions, mixed with different dependence structures, are simulated to compare its results with those of other procedures. For this purpose, a Monte Carlo study is carried out. Three different interesting dependence scenarios are introduced, simulating them under the linear regression model structure given by (\ref{linear_regre}). We consider $\beta$ as a sparse vector of length $p$ with only $s<p$ values not equal zero and $X\in\mathbb{R}_{n\times p}$, where $n$ is the sample size. It is assumed that $\varepsilon\in N_n(0,\sigma^2I_n)$, so once a value for $\sigma^2$ is provided, we can obtain $Y$. We fix $p=100$ and choose $\sigma^2$ verifying that the percentage of explained deviance is explicitly the $90\%$. Calculation of this parameter is collected in Section 1 of the Supplementary material. Then, to guarantee the conditions introduced in Sections \ref{consistency}, \ref{false_discoveries} and \ref{lambda_selection},  it is needed that $n>4.61s$ as we saw in (\ref{consistet_LASSO_P}), $\inf |\beta_j| >> 2.15\sqrt{s/ n}$ for $j\in S$ as in (\ref{beta_size}) and $\lambda \sim 2.15\sigma\sqrt{1/n}$. To test their performance under these conditions and when they are violated, we consider different combinations of parameters values taking $n=25,50,100,200,400$ and $s=10,15,20$. A study of when these conditions hold is showed in Section 2 of the Supplementary material. In every simulation, we count the number of covariates correctly selected ($|\hat{S}\cap S|$) as well as the noisy ones ($|\hat{S}\setminus S|$). Besides, we measure the prediction power of the algorithm by means of the percentage of explained deviance ($\% Dev$) and the mean squared error ($MSE$). This last gives us an idea about the bias produced by the LASSO (see Section \ref{biased_estimator}). We repeat this procedure a number $M=500$ of times and compute on average its results.

\begin{itemize}
	\item{\textbf{Scenario 1} (\textit{Orthogonal design}). Only the first $s$ values are not equal zero for $\beta_j$ with $j=1,\dots,s$ and $p>s>0$, $\beta_1=\dots=\beta_{s}=1.25$, while $\beta_j=0$ for all $j=s+1,\dots,p$. $X$ is simulated as a $N_n(0,I_p)$.}
	
	\item{\textbf{Scenario 2} (\textit{Dependence by blocks}). The vector $\beta$ has the first $s<p$ components not null, of the form $\beta_1=\dots=\beta_{s}=1$ and $\beta_j=0$ for the rest. $X$ is simulated as a $N_n(0,\Sigma)$, where $\sigma_{jj}=1$ and $\sigma_{jk}=cov(X_j, X_k)=0$ for all pairs $(j,k)$ except if $mod_{10}(j)=mod_{10}(k)$, in that case $\sigma_{jk}=\rho$, taking $\rho=0.5,0.9$.}
	
	\item{\textbf{Scenario 3} (Toeplitz covariance). Again, only $s$ ($p>s>0$) covariates are important, simulating $X$ as a $N_n(0,\Sigma)$ and assuming $\beta_j=0.5$ in the places where $\beta\not=0$. In this case, $\sigma_{jk}=\rho^{|j-k|}$ for $j,k=1,\dots,p$ and $\rho=0.5,0.9$. Now, we analyze two different dependence structures varying the location of the $s$ relevant covariates:
		\begin{itemize}
			\item{\textbf{Scenario 3.a}:  we assume that the relevant covariates are the first $s=15$.}
			\item{\textbf{Scenario 3.b}: consider $s=10$ relevant variables placed every 10 sites, which means that only the $\beta_1,\beta_{11},\beta_{21},\dots,\beta_{91}$ terms of $\beta$ are not null.}
		\end{itemize}}
	
\end{itemize}

The first choice, the orthogonal design of Scenario 1, is selected as the best possible framework. This verifies the consistency conditions for values of $n$ large enough and avoids the confusion phenomenon given that there are not correlated covariates.

In contrast, to assess how the LASSO behaves in case of different dependence structures, Scenario 2 and Scenario 3 are proposed. In the dependence by blocks context (Scenario 2), we force the design to have a dependence structure where the covariates are correlated ten by ten. As a result, we induce a more challenging scenario for the LASSO, in which the algorithm has to overcome a fuzzy signal produced by irrelevant covariates. Different magnitudes of dependence are considered in Scenario 2 with $\rho=0.5$ and $\rho=0.9$ to test the effect of the confusion phenomenon. As a result, different sizes of $n$ are needed in terms of $s$ to guarantee the proper behavior of the LASSO. This scenario has been studied in other works, like in \cite{Meinshausen2010}. 

Eventually, the LASSO performance is tested in a scenario were all the covariates are correlated: the Toeplitz covariance structure (Scenario 3). This mimics a time series dependence pattern. This is an example where the irrepresentable condition holds but the algorithm suffers from highly correlated relations between the true set of covariates and unimportant ones. This framework has been studied for the LASSO case, see for example \cite{Meinshausen2010} or \cite{buhlmann2011statistics}. Because the distance between covariates is relevant to establish their dependence, we study two different frameworks. In the first scenario (Scenario 3.a) the important covariates are highly correlated among them and little with the rest. Particularly, there are only notable confusing correlations in the case of the last variables of $S=\{1,\dots,15\}$ with their noisy neighbors. Here, the LASSO is only able to recover $S$ in the $n=400$ case. In contrast, in the Scenario 3.b, the important covariates are markedly correlated with unimportant ones, contributing to magnify the spurious correlations phenomenon. For this scenario we need a size of $n=200,400$.

We start testing the performance of the standard LASSO using the library \texttt{glmnet} (\cite{Friedman2010}) implemented in R (\cite{R}). This uses K-fold cross-validation to select the $\lambda$ parameter. As it was explained in Section \ref{lambda_selection}, this is one of the most popular ways of estimating $\lambda$. In order to be capable of comparing different models and following recommendations of the existing literature, we have fixed $K=10$ for all simulations. Besides, we work with the response $y$ centered and with the matrix $\mathbf{X}$ standardized by columns. This last would not be really necessary in these frameworks because the covariates are all in the same scale. However, this is done to keep the usual implementation of LASSO type algorithms in practice. We apply a first screening step and then we adjust a linear regression model with the selected covariates. This scheme is also following for the rest of procedures.

There are other faster algorithms available in R, such as the famous LARS procedure (\cite{efron2004least}, \cite{lars_R}). However, we decided to make use of the \texttt{glmnet} library due to its easy implementation and interpretation, as well as its simple adaptation to other derivatives of the LASSO we test in this document.

\subsection{Performance of the LASSO in practice}\label{Performance LASSO}

In order to test the inconveniences of the LASSO when there exists dependence among covariates, a complete simulation study is carried out. For this purpose, we make use of the simulation scenarios introduced above. Complete results are displayed in the Section 4 of the Supplementary material.


In the orthogonal design of Scenario 1, we would expect the LASSO to recover the whole set of important covariates and not to add too much noise into the model for a large enough value of $n$. However, we have observed different results.

\begin{figure}[htb]\centering
	\includegraphics[width=\linewidth]{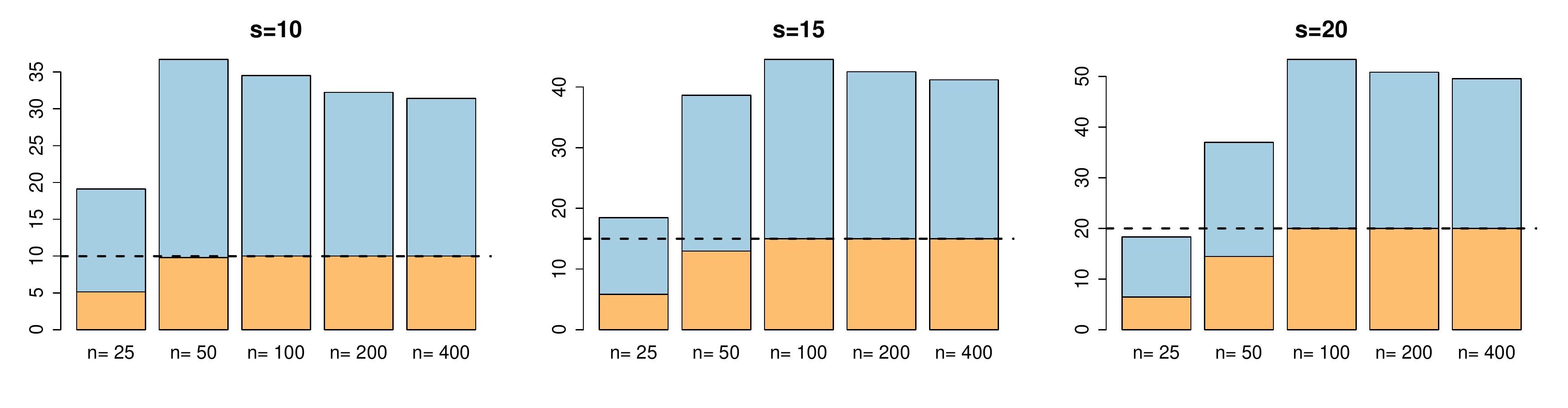}
	\caption{\label{porcentajes_M1} Number of important covariates (\textcolor{orange}{orange} area) versus noisy ones (\textcolor{bluegray}{blue} area) selected by the LASSO in Scenario 1. The dashed line marks the $s$ value.}
\end{figure}

Firstly, we can appreciate that it does not really matter the number of relevant covariates considered ($s=10,15,20$) in relation with the capability of recovering this set. It is because the algorithm only includes the complete set under the $n\geq p$ framework except for the $s=10$ scenario taking $n=50$. See this fact in Figure \ref{porcentajes_M1}. It can be easily explained in terms of the consistence requirements of the LASSO given in (\ref{consistet_LASSO_P}). Besides, although we are under orthogonal design assumption, this includes a lot of noisy variables in the model. What is shocking is the fact that the number of irrelevant covariates selected is always larger than the important ones.  This exemplifies the existing trade-off between FDP and TPP introduced in (\ref{FDP_TPP}) as well as that both quantities can not be simultaneously low.

\begin{table}[htb] 
	\centering
	\small 
	\begin{tabular}{ccccccc}
		\cline{2-7}
		\rule{0pt}{0.4cm} & \multicolumn{2}{c}{$\mathbf{s=10}$} & \multicolumn{2}{c}{$\mathbf{s=15}$} & \multicolumn{2}{c}{$\mathbf{s=20}$} \\	
		\cmidrule(r){2-3} \cmidrule(lr){4-5} \cmidrule(l){6-7}
		\rule{0pt}{0.4cm} & \multicolumn{1}{c}{\textbf{MSE} \scriptsize$(1.736)$} &  \textbf{\% Dev}  & \multicolumn{1}{c}{\textbf{MSE} \scriptsize$(2.604)$} &  \textbf{\% Dev} & \multicolumn{1}{c}{\textbf{MSE} \scriptsize$(3.472)$} &  \textbf{\% Dev} \\
		\hline
		\multicolumn{1}{c}{\rule{0pt}{0.4cm} $n=50$} & 0.18 & 0.989 & 0.517 & 0.977 & 1.668 & 0.947 \\
		\multicolumn{1}{c}{$n=100$} & 0.701 & 0.959 & 0.846 & 0.967 & 0.916 & 0.973 \\
		\multicolumn{1}{c}{$n=200$} & 1.164 & 0.932 & 1.624 & 0.936 & 2.060 & 0.94  \\ 
		\hline
	\end{tabular}
	\caption{Summary of the LASSO results for Scenario 1. The oracle value for the deviance is $0.9$ and those for the MSE are in brackets. }
	\label{compare_LASSO_M1}
\end{table}

In second place, we notice that this procedure clearly overestimates its results. This obtains values for the MSE and percentage of explained deviance less and greater, respectively, of the oracle ones (see values in brackets in Table \ref{compare_LASSO_M1}). In conclusion, with this toy example we can illustrate how the LASSO procedure performs very poorly and present important limitations even in an independence framework.


Next, we analyze the results of the dependence by blocks context. In case of dependence, it is expected for a ``smart'' algorithm to be capable of selecting a portion of relevant covariates and explaining the remaining ones making use of the existing correlation structure. The subset of $S$ which is really necessary to explain this type of models is denote as ``effective covariates''. These can be calculated measuring how many are necessary to explain a certain percentage of $\Sigma_S$ variability, being $\Sigma_S$ the submatrix of $\Sigma$ considering the elements of $S$. This number is inversely proportional to the dependence strength. For example, to explain the $90-95\%$, we found that for the Scenario 2 with $\rho=0.5$ there are needed about $12-14$ covariates taking $s=15$ and about $16-18$ for the case of $s=20$. In contrast, only $10$ are necessary in Scenario 2 with $\rho=0.9$. Complete calculation for the different simulation scenarios is displayed in Section 3 of the Supplementary material. Again, the LASSO presents some difficulties for an efficient recovery. 

\begin{figure}[htb]\centering
	\includegraphics[width=\linewidth]{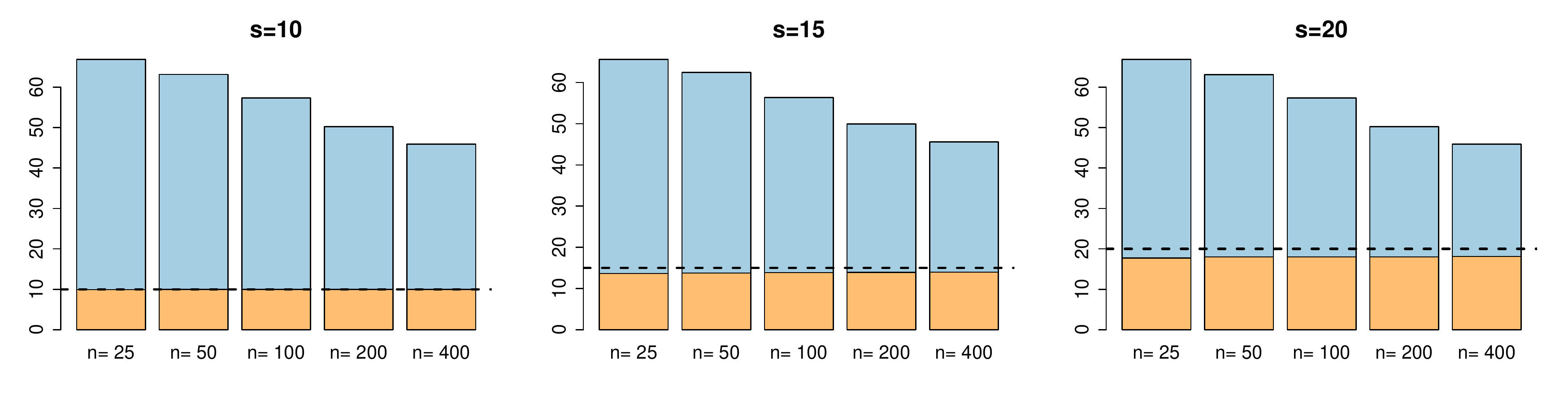}
	\caption{\label{porcentajes_M2_rho_0_5} Number of important covariates (\textcolor{orange}{orange} area) versus noisy ones (\textcolor{bluegray}{blue} area) selected by the LASSO in Scenario 2 with $\rho=0.5$. The dashed line marks the $s$ value.}
\end{figure}

\begin{table}[htb] 
	\centering
	\small 
	\begin{tabular}{ccccccc}
		\cline{2-7}
		\rule{0pt}{0.4cm} & \multicolumn{2}{c}{$\mathbf{s=10}$} & \multicolumn{2}{c}{$\mathbf{s=15}$} & \multicolumn{2}{c}{$\mathbf{s=20}$} \\	
		\cmidrule(r){2-3} \cmidrule(lr){4-5} \cmidrule(l){6-7}
		\rule{0pt}{0.4cm} & \multicolumn{1}{c}{\textbf{MSE} \scriptsize$(0.556)$} &  \textbf{\% Dev} & \multicolumn{1}{c}{\textbf{MSE} \scriptsize$(1.389)$} &  \textbf{\% Dev} & \multicolumn{1}{c}{\textbf{MSE} \scriptsize$(2.222)$} &  \textbf{\% Dev} \\ 
		\hline
		\multicolumn{1}{c}{\rule{0pt}{0.4cm} $n=50$} & 0.438 & 0.956 & 1.095 & 0.956 & 1.752 & 0.956 \\
		\multicolumn{1}{c}{$n=100$} & 0.495 & 0.951 & 1.238 & 0.951 & 1.981 & 0.951 \\
		\multicolumn{1}{c}{$n=200$} & 0.523 & 0.951 & 1.307 & 0.950 & 2.091 & 0.951  \\ 
		\hline
	\end{tabular}
	\caption{Summary of the LASSO results for Scenario 2 with $\rho=0.5$. The oracle value for the deviance is $0.9$ and those for the MSE are in brackets. }
	\label{compare_LASSO_M2_0.5}
\end{table}

\begin{figure}[htb]\centering
	\includegraphics[width=\linewidth]{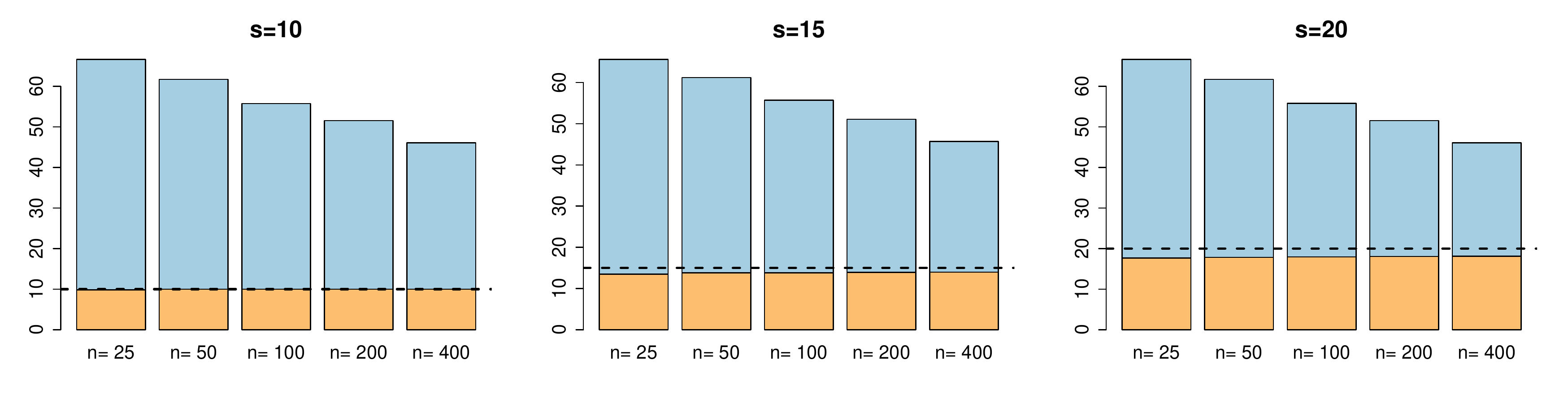}
	\caption{\label{porcentajes_M2_rho_0_9} Number of important covariates (\textcolor{orange}{orange} area) versus noisy ones (\textcolor{bluegray}{blue} area) selected by the LASSO in Scenario 2 with $\rho=0.9$. The dashed line marks the $s$ value.}
\end{figure}

A summary of the results for the Scenario 2 with $\rho=0.5$ is displayed in Table \ref{compare_LASSO_M2_0.5} and Figure \ref{porcentajes_M2_rho_0_5}, while for the Scenario 2 with $\rho=0.9$ is showed in Table \ref{compare_LASSO_M2_0.9} and Figure \ref{porcentajes_M2_rho_0_9}. If we consider only $s=10$ important explanatory variables, its behavior is quite similar to the Scenario 1. Besides, in both scenarios with $s=10$, LASSO almost recovers the complete set $S$, even for $n=25$ although its proper recovery is guaranteed from $n=50$. However, more noise is included in this last.

\begin{table}[htb] 
	\centering
	\small 
	\begin{tabular}{ccccccc}
		\cline{2-7}
		\rule{0pt}{0.4cm} & \multicolumn{2}{c}{$\mathbf{s=10}$} & \multicolumn{2}{c}{$\mathbf{s=15}$} & \multicolumn{2}{c}{$\mathbf{s=20}$} \\	
		\cmidrule(r){2-3} \cmidrule(lr){4-5} \cmidrule(l){6-7}
		\rule{0pt}{0.4cm} & \multicolumn{1}{c}{\textbf{MSE} \scriptsize$(1)$} &  \textbf{\% Dev} & \multicolumn{1}{c}{\textbf{MSE} \scriptsize$(2.5)$} &  \textbf{\% Dev} & \multicolumn{1}{c}{\textbf{MSE} \scriptsize$(4)$} &  \textbf{\% Dev} \\ 
		\hline
		\multicolumn{1}{c}{\rule{0pt}{0.4cm} $n=50$} & 0.784 & 0.926 & 1.96 & 0.925 &3.137 &  0.926 \\
		\multicolumn{1}{c}{$n=100$} & 0.888 & 0.918 & 2.22 & 0.918 & 3.551 & 0.918 \\
		\multicolumn{1}{c}{$n=200$} & 0.939 & 0.913 & 2.347 & 0.913 & 3.756 & 0.913  \\ 
		\hline
	\end{tabular}
	\caption{Summary of the LASSO results for Scenario 2 with $\rho=0.9$. The oracle value for the deviance is $0.9$ and those for the MSE are in brackets. }
	\label{compare_LASSO_M2_0.9}
\end{table}

In contrast, the situation is different if we simulate with $s=15$ or $s=20$ relevant covariates. Then, the LASSO does not tend to recover the covariates of $S$, not even for values of $n$ verifying $n\geq p$ as well as conditions (\ref{consistet_LASSO_P}) and (\ref{beta_size}). See Section 2 of the Supplementary material for more information. However, this selects more than the effective number of covariates. It seems the LASSO tries to recover the set $S$ but, due to the presence of spurious correlations, this chooses randomly between two highly correlated important covariates. 
It can be appreciated in Figures \ref{barplots_M2_25_rho_5}-\ref{barplots_M2_clases_rho_9} 
of the Appendix that the $10$ first covariates are selected with high probability, near $1$, but due to the confusion phenomenon some of them are interchanged by a representative one. The following $s-10$ relevant variables have a lower selection rate and there are some irrelevant ones selected a larger number of times, adding  quite noise to the model. This inconvenient seems not to be overcome increasing the number of samples $n$. Again, the LASSO keeps overestimating its results as we can see by the percentage of explained deviance and the MSE.


\begin{figure}[htb]\centering
	\includegraphics[width=\linewidth]{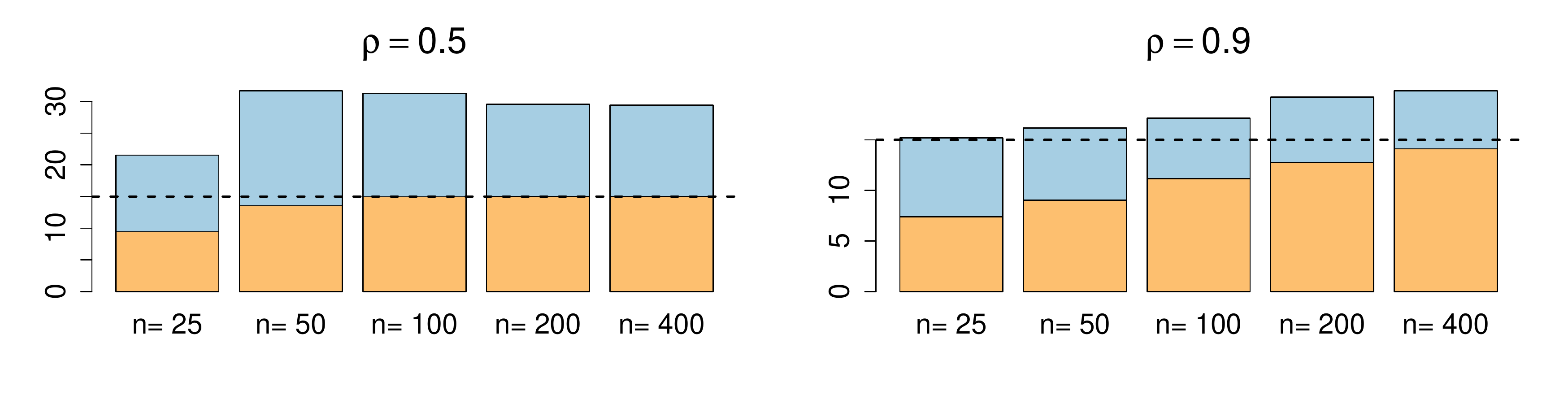}
	\caption{\label{porcentajes_M3_s_15} Number of important covariates (\textcolor{orange}{orange} area) versus noisy ones (\textcolor{bluegray}{blue} area) selected by the LASSO in Scenario 3.a. The dashed line marks the $s$ value.}
\end{figure}

\begin{table}[htb] 
	\centering
	\small 
	\begin{tabular}{ccccc}
		\cline{2-5}
		\rule{0pt}{0.4cm} & \multicolumn{2}{c}{$\boldsymbol{\rho=0.5}$} & \multicolumn{2}{c}{$\boldsymbol{\rho=0.9}$}  \\	
		\cmidrule(lr){2-3} \cmidrule(lr){4-5} 
		\rule{0pt}{0.4cm} & \multicolumn{1}{c}{\textbf{MSE} \scriptsize$(1.139)$} &  \textbf{\% Dev} & \multicolumn{1}{c}{\textbf{MSE} \scriptsize$(3.807)$} &  \textbf{\% Dev} \\  
		\hline
		\multicolumn{1}{c}{\rule{0pt}{0.4cm} $n=50$} & 0.19 & 0.983 & 1.894 & 0.950  \\
		\multicolumn{1}{c}{$n=100$} & 0.546 & 0.951 & 2.815 & 0.928  \\
		\multicolumn{1}{c}{$n=200$} & 0.825 & 0.927 & 3.302 & 0.916  \\ 
		\hline
	\end{tabular}
	\caption{Summary of the LASSO results for Scenario 3.a. The oracle value for the deviance is $0.9$ and those for the MSE are in brackets. }
	\label{compare_LASSO_M3_s1}
\end{table}

\begin{figure}[htb]\centering
	\includegraphics[width=\linewidth]{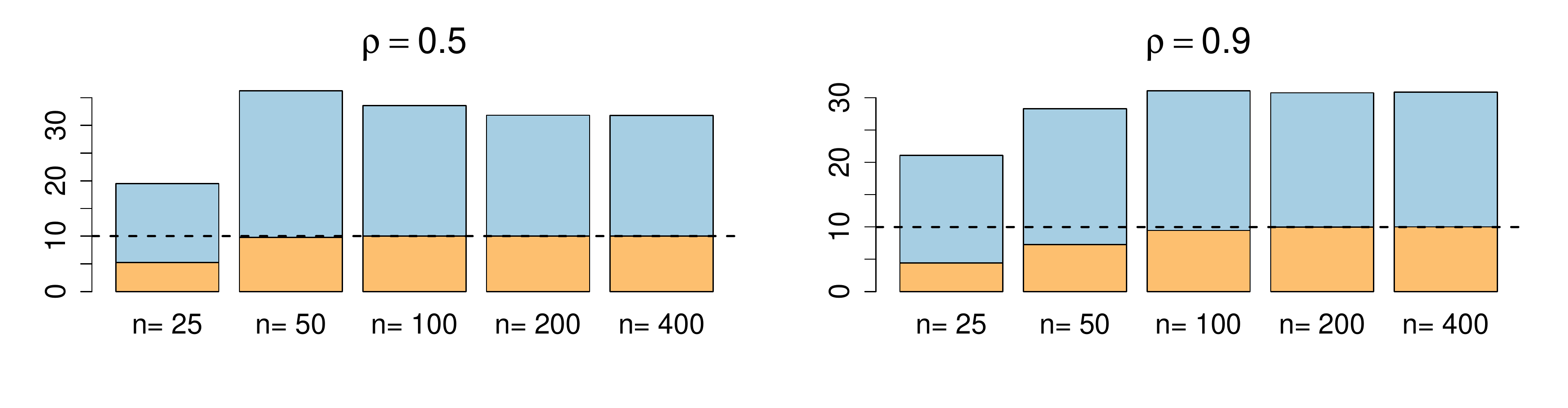}
	\caption{\label{porcentajes_M3_s_10}  Number of important covariates (\textcolor{orange}{orange} area) versus noisy ones (\textcolor{bluegray}{blue} area) selected by the LASSO in Scenario 3.b. The dashed line marks the $s$ value.}
\end{figure}

Finally, we study the results of the Toeplitz covariance structure by means of the Scenario 3.a, where the relevant covariates are the first $s=15$ (Table \ref{compare_LASSO_M3_s1} and Figure \ref{porcentajes_M3_s_15}), and the Scenario 3.b, where there are only $s=10$ important variables placed every 10 sites (Table \ref{compare_LASSO_M3_s2} and Figure \ref{porcentajes_M3_s_10}). 

 Interpreting their results, we see that the LASSO procedure recovers the important set of covariates for $\rho=0.5$, taking a value of $n=100,200,400$ verifying the consistent condition, in both cases. Nevertheless, this exceeds the number of efficient covariates in Scenario 3.a  for $\rho=0.9$ and in the two cases of Scenario 3.b, because with $10$ covariates it is explained the $98\%$ of variability. Moreover, this algorithm returns to include many pointless covariates in the model and overestimates the prediction accuracy.

\begin{table}[h!] 
	\centering
	\small 
	\begin{tabular}{ccccc}
		\cline{2-5}
		\rule{0pt}{0.4cm} & \multicolumn{2}{c}{$\boldsymbol{\rho=0.5}$} & \multicolumn{2}{c}{$\boldsymbol{\rho=0.9}$} \\	
		\cmidrule(lr){2-3} \cmidrule(lr){4-5}  
		\rule{0pt}{0.4cm} & \multicolumn{1}{c}{\textbf{MSE} \scriptsize$(0.278)$} &  \textbf{\% Dev} & \multicolumn{1}{c}{\textbf{MSE} \scriptsize$(0.53)$} &  \textbf{\% Dev} \\ 
		\hline
		\multicolumn{1}{c}{\rule{0pt}{0.4cm} $n=50$} & 0.034 & 0.987 & 0.147 & 0.971 \\
		\multicolumn{1}{c}{$n=100$} & 0.123 & 0.955 & 0.309 & 0.94 \\
		\multicolumn{1}{c}{$n=200$} & 0.195 & 0.929 & 0.417 & 0.920  \\ 
		\hline
	\end{tabular}
	\caption{Summary of the LASSO results for Scenario 3.b. The oracle value for the deviance is $0.9$ and those for the MSE are in brackets. }
	\label{compare_LASSO_M3_s2}
\end{table}


\section{Evolution of the LASSO in the last years and alternatives}\label{other}	

Once the LASSO and its inconveniences have been displayed, we want to compare its performance with other approaches. Then, we test different methodologies designed to select the relevant information and adjust a regression model in the $p>n$ framework. Nevertheless, it is impossible to include all the existing algorithms here. Instead, we attempted to collect the most relevant ones, providing a summary of the most used methodologies nowadays.

Methods proposed to alleviate the limitations of the LASSO algorithm are based in a wide range of different philosophies. Some of them opt to add a second selection step after solving the LASSO problem, such as the relaxed LASSO (\cite{meinshausen2007relaxed}) or thresholded LASSO (\cite{Zhou2010}, \cite{Geer2011}), while other alternatives are focused on giving different weights to the covariates proportional to their importance, as the adaptive LASSO (\cite{Zou2006}, \cite{huang2008adaptive}, \cite{Geer2011}). Others pay attention to the group structure of the sparse vector $\beta$ when this exists, like the grouped LASSO procedure (\cite{yuan2006model}) or the fused LASSO (\cite{tibshirani2005sparsity}). 

The resampling or iterative procedures are other approaches which make use of subsampling or computational power, algorithms like boLASSO (\cite{Bach2008}), stability selection with randomized LASSO (\cite{Meinshausen2010}), the random LASSO (\cite{wang2011random}), the scaled LASSO (\cite{sun2012}) or the combination of classic estimators with variable selection diagnostics measures (\cite{nan2014variable}), among others, are based on this idea. Furthermore, more recent techniques like the Knockoff filter (\cite{Barber2015}, \cite{candes2018panning})  or SLOPE (\cite{Bogdan2015}) have been introduced to control some measures of the type I error. One drawback of the famous Knockoff filter is that, to the best of our knowledge, this is not yet available for the $p>n$ case.

There are other alternatives, which modify the constraints of the LASSO problem (\ref{bLASSO}) so as to achieve better estimators of $\beta$, like the Elastic Net (\cite{Zou2005}), the Dantzig selector (\cite{candes2007dantzig},  \cite{bickel2009simultaneous}) or the square root LASSO (\cite{belloni2011square}). Moreover, other different alternatives have been developed recently, such as the Elem-OLS Estimator (\cite{yang2014elementary}), the LASSO-Zero (\cite{descloux2018model}) or the horseshoe (\cite{bhadra2019LASSO}), adding new ideas to the previous list.

Quoted \cite{descloux2018model}: although differing in their purposes and performance, the general idea underlying these procedures remains the same, namely to avoid overfitting by finding a trade-off between the fit $y-X\beta$ and some measure of the model complexity.

Along the many papers, we have found that a modest classification of the different proposals can be done, although in this classification some of the procedures does not only fit in a single class. These categories are

\begin{itemize}
	\item{\textbf{Weighted LASSO:} weighted versions of the LASSO algorithm with a suitable selection of the weights. They are proposed to attach the particular importance of each covariate in the estimation process. Besides, joint with iteration, this modification allows a reduction of the bias. }
	\item{\textbf{Resampling LASSO procedures:} mix of the typical LASSO adjustment with resampling procedures, based, for example, on randomization in the selection process of covariates so as to reduce unavoidable random noise.}
	\item{\textbf{Thresholded versions of the LASSO:} a second thresholding step in the covariates selection is implemented in order to reduce the irrelevant ones.}
	\item{\textbf{Alternatives to the LASSO:} procedures with different nature and aims designed to solve the LASSO drawbacks.}
\end{itemize}

This extensive list of procedures makes noticeable the impact the LASSO has nowadays. A brief summary is displayed in Table \ref{table_other}.

Owing to the computational cost required for the resampling LASSO procedures, such as the boLASSO of \cite{Bach2008} or the random LASSO algorithm (\cite{wang2011random}), these algorithms are too slow. Even for small values of $p$, we found that the computational costs were high. For this reason, they are excluded for the comparative analysis studio. The LASSO-Zero technique of \cite{descloux2018model} suffers from the same issue, so it is excluded too.

Other problem springs up for the thresholded versions of the LASSO. In this case, we noticed that the complexity of finding a correct threshold is similar to the one of obtaining the optimal value of $\lambda$ for the LASSO adjustment. In both cases, we would need to know in advance the dispersion of the error $\sigma^2$, which is usually impossible in practice. Then, procedures as the thresholded LASSO algorithm of \cite{Zhou2010} are shut out in order to not add more complications to the adjustment.

\begin{figure}[htb]\centering
	\includegraphics[width=\linewidth]{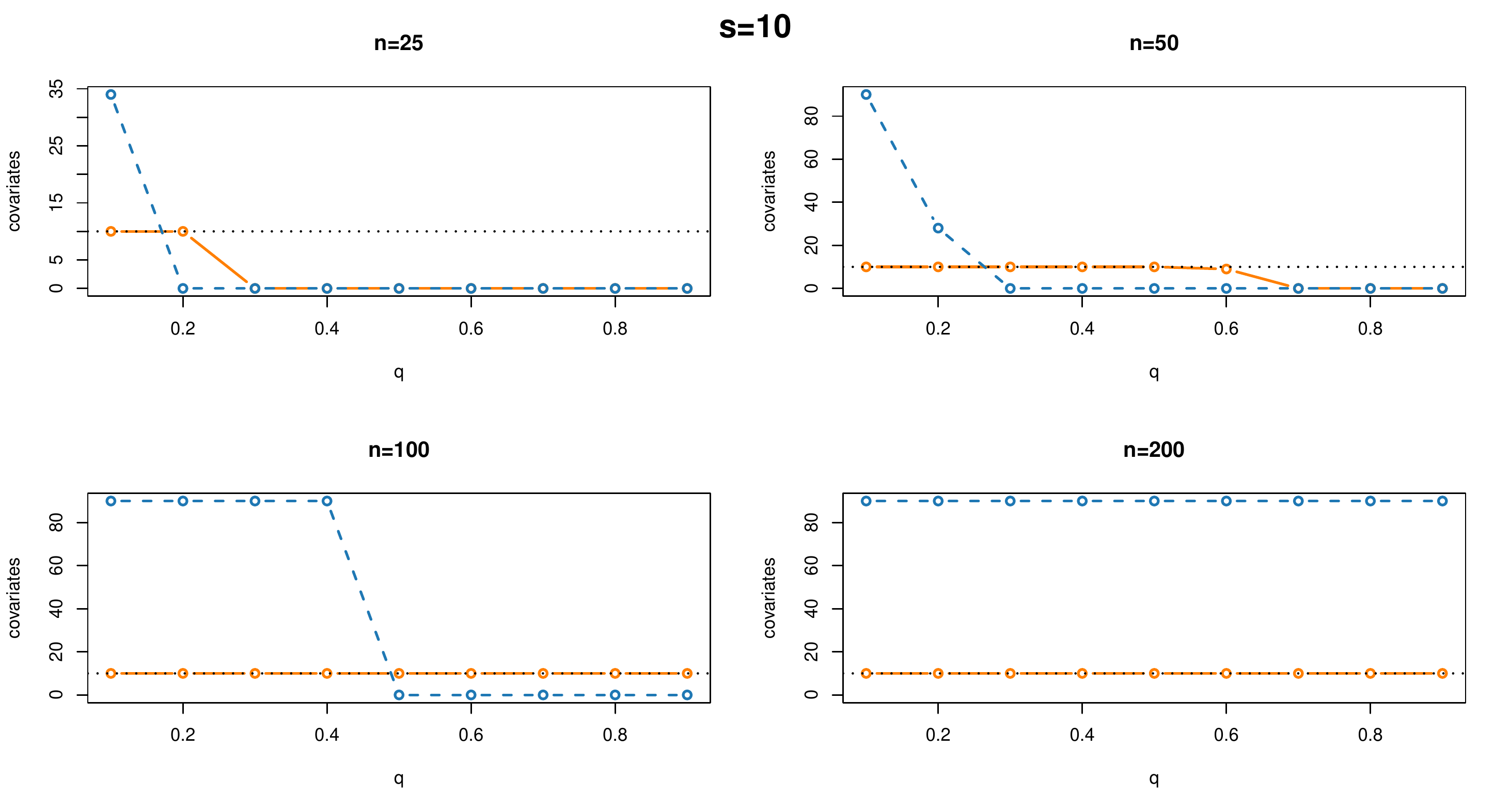}
	\caption{\label{threshold_s_10} Number of important covariates (\textcolor{orange}{orange}) versus noisy ones (\textcolor{bluegray}{blue}) selected by the stability selection algorithm applying LASSO and taking different thresholds $q$, for $s=10$ in the Scenario 1. The horizontal dashed lines are the true relevant covariates in each of them.}
\end{figure}

One method in the middle of both groups is the stability selection procedure proposed by \cite{Meinshausen2010}. This methodology pays attention to the probability of each covariate to be selected. Only the covariates with probability greater than a fixed threshold are added to the final model. We have observe in practice that a proper choice of the threshold value seems to depend on the sample size considered, $n$, as well as the sparsity of the vector $\beta$. An example is showed simulating the stability selection procedure using the standard LASSO in Scenario 1 with $s=10$ (Figure \ref{threshold_s_10}). Eventually, it is interesting to take into account that it seems no possible to guarantee consistency for any thresholding value in case that $n>p$. For this reason, this approach is not included in the comparison neither.

\begin{longtable}{|cc|c|}
	\hline
	\rule{0pt}{0.5cm} \textbf{PROBLEM FORMULATION} &  & \textbf{PROS} \rule[-0.3cm]{0pt}{0pt}\\
	\hline
	\endfirsthead
	
	\hline
	\rule{0pt}{0.5cm} \textbf{PROBLEM FORMULATION} & & \textbf{PROS} \rule[-0.3cm]{0pt}{0pt}\\
	\hline
	\endhead
	
	\rule{0pt}{0.5cm} {\LARGE \textcolor{darktangerine}{$\mathbf{\ast}$}} \textbf{Best subset selection} -- \cite{beale1967discarding}, & & \multirow{4}*{\textit{Better selection}} \\
	 \hspace{5.5cm} \cite{hocking1967selection} & &  \\
	\rule{0pt}{0.5cm} {\small $ \stackbin[\beta]{}{\min} \left\lbrace \sum_{i=1}^{n}\left(y_i-\sum_{j=1}^{p}x_{ij}\beta_j\right)^2+ \lambda\sum_{j=1}^{p} \mathbf{1}_{\beta_j \not=0} \right\rbrace$} & \textcolor{red}{\texttimes} \rule[-0.6cm]{0pt}{0pt} & \\
	\hline 
	\rule{0pt}{0.5cm} \textbf{LASSO} -- \cite{Tibshirani1996} & & \multirow{4}*{\textit{--}} \\
	\rule{0pt}{0.7cm} {\small $\stackbin[\beta]{}{\min} \left\lbrace \sum_{i=1}^{n}\left(y_i-\sum_{j=1}^{p}x_{ij}\beta_j\right)^2+ \lambda\sum_{j=1}^{p}|{\beta_j}| \right\rbrace$} & \textcolor{darkpastelgreen}{$\checkmark$}  \rule[-0.6cm]{0pt}{0pt} & \\
	\hline
	\rule{0pt}{0.5cm} {\LARGE \textcolor{darktangerine}{$\mathbf{\ast}$}} \textbf{SCAD} --  \cite{fan1997comments} & & \multirow{8}*{\textit{Better selection}} \\
	\rule{0pt}{0.7cm} {\small$\stackbin[\beta]{}{\min} \left\lbrace \sum_{i=1}^{n}\left(y_i-\sum_{j=1}^{p}x_{ij}\beta_j\right)^2+ p_{\lambda}(\beta) \right\rbrace$} &
	\multirow{5}*{\textcolor{red}{\texttimes}} & \multirow{7}*{\textit{Bias reduction}} \\
	\rule{0pt}{1.5cm} with {\small $p_{\lambda}(\beta)= \left\lbrace
	\begin{aligned}
	& \lambda |\beta|, \quad &\text{if} \; |\beta|\leq \lambda, \\
	&\frac{2a\lambda |\beta|-\beta^2-\lambda^2}{2(a-1)},  \quad &\text{if} \; \lambda < |\beta| \leq a\lambda \quad (a>2) \\
	& \frac{\lambda^2 (a+1)}{2}, \quad &\text{otherwise}.
	\end{aligned} 
	\right. $}  &  &  \rule[-1.3cm]{0pt}{0pt} \\
	\hline
	\rule{0pt}{0.5cm} \textbf{Basis Pursuit Denoising} -- \cite{chen2001atomic} &  &  \\
	\rule{0pt}{0.4cm} {\small $\stackbin[\beta]{}{\min}  \|\beta\|_1 \quad \text{subject to} \; \| y-X\beta\|_2\leq \theta$} & \multirow{1}*{\textcolor{red}{\texttimes}} & \multirow{1}*{\textit{--}} \rule[-0.4cm]{0pt}{0pt} \\
	\hline
	\rule{0pt}{0.6cm} {\LARGE \textcolor{darktangerine}{$\mathbf{\ast}$}} \textbf{Elastic Net} -- \cite{Zou2005} &  & \multirow{2}{3cm}{\centering \textit{Better prediction\\  \vspace{0.2cm} Possible selection of more than $n$ covariates ($p>n$) }} \\
	\rule{0pt}{0.4cm} {\small $\stackbin[\beta]{}{\min} \left\lbrace \sum_{i=1}^{n}\left(y_i-\sum_{j=1}^{p}x_{ij}\beta_j\right)^2+ \lambda\sum_{j=1}^{p}\left(\alpha|{\beta_j}|+(1-\alpha)\beta_j^2\right) \right\rbrace$} & \multirow{2}*{\textcolor{darkpastelgreen}{$\checkmark$}} & \\
    {\small $\text{with}\,\alpha\in(0,1)$} & \rule[-0.4cm]{0pt}{0pt} &  \\
	\hline
	\rule{0pt}{0.5cm} {\LARGE \textcolor{darktangerine}{$\mathbf{\ast}$}} \textbf{Fused LASSO} --  \cite{tibshirani2005sparsity} & & \multirow{3}*{\textit{Ordered structure}}\\
	\rule{0pt}{0.5cm} {\small $ \hspace{-0.5cm} \stackbin[\beta]{}{\min} \left\lbrace \sum_{i=1}^{n}\left(y_i-\sum_{j=1}^{p}x_{ij}\beta_j\right)^2 \right.$} &  & \\
	 {\small $ \hspace{3cm} \left. + \lambda_1\sum_{j=1}^{p}|{\beta_j}| + \lambda_2\sum_{j=2}^{p}|\beta_j-\beta_{j-1}| \right\rbrace$} & \textcolor{darkpastelgreen}{$\checkmark$}  \rule[-0.6cm]{0pt}{0pt} & \\
	\hline
	\rule{0pt}{0.5cm} {\LARGE \textcolor{byzantine}{$\mathbf{\bullet}$}} \textbf{Adaptive LASSO} -- \cite{Zou2006} & & \\
	\rule{0pt}{0.5cm} {\small $\stackbin[\beta]{}{\min} \left\lbrace \sum_{i=1}^{n}\left(y_i-\sum_{j=1}^{p}x_{ij}\beta_j\right)^2+ \lambda\sum_{j=1}^{p} w_j|{\beta_j}| \right\rbrace$} & \multirow{3}*{\textcolor{darkpastelgreen}{$\checkmark$}} & \multirow{2}*{\textit{Better selection}} \\
	\rule{0pt}{0.5cm} {\footnotesize (taking $w_j=1/|\hat{\beta}^{RR}_j|^q$ where $\hat{\beta}^{RR}$ is the ridge estimator } & & \textit{Bias reduction} \\
	{\footnotesize (\cite{hoerl1970ridge}) and $q\geq1$)}& \rule[-0.3cm]{0pt}{0pt} & \\
	\hline
	\rule{0pt}{0.5cm} {\LARGE \textcolor{darktangerine}{$\mathbf{\ast}$}} \textbf{Group LASSO} -- \cite{yuan2006model} & & \\
	\rule{0pt}{0.6cm} {\small $\stackbin[\beta]{}{\min} \left\lbrace \sum_{i=1}^{n}\left(y_i-\sum_{k=1}^{K}X_{k}\beta_k\right)^2+ \lambda\sum_{k=1}^{K}\|{\beta_k}\|_{Z_k}  \right\rbrace$} & \multirow{2}*{\textcolor{darkpastelgreen}{$\checkmark$}} & \multirow{2}*{\textit{Group structure}} \\
    {\small $ \text{with} \; \| w\|_{Z_k}=(w^\top Z_k w)^{1/2}$} &   & \\
	\rule{0pt}{0.5cm} {\footnotesize ($Z_k$ are kernel matrices of the functional space induced by the $k$th factor)}& \rule[-0.3cm]{0pt}{0pt} & \\
	\hline
	\pagebreak
	\rule{0pt}{0.5cm} {\LARGE \textcolor{darktangerine}{$\mathbf{\ast}$}} \textbf{Dantzig selector} -- \cite{candes2007dantzig} & &  \multirow{5}{3cm}{\centering \textit{Consistent to orthogonal transformations}}\\
	\rule{0pt}{0.5cm} {\small $\stackbin[\beta]{}{\min}  \|\beta\|_1 \quad \text{subject to} \;  \| X^\top r \|_{\infty} \leq \lambda_p \cdot \sigma $} & \multirow{2}*{\textcolor{red}{\texttimes}}  & \\
	\rule{0pt}{0.5cm} {\footnotesize (with $\| X^\top r \|_{\infty}:= \stackbin[1\leq j \leq p]{}{\sup} |(X^\top r)_j| $ and $r=y-X \beta$)} & \rule[-0.5cm]{0pt}{0pt} & \\
	\hline
	\rule{0pt}{0.1cm} & & \multirow{5}{3cm}{\centering \textit{Faster convergence rates\\ \vspace{0.2cm} More accurate predictions}}\\
	 {\LARGE \textcolor{darktangerine}{$\mathbf{\ast}$}} \textbf{Relaxed LASSO} -- \cite{meinshausen2007relaxed} & & \\
	\rule{0pt}{0.5cm} {\small $\stackbin[\beta]{}{\min} \left\lbrace n^{-1}\sum_{i=1}^{n}\left( y_i-x_i^\top \{\beta \cdot \mathbf{1}_{\mathcal{M}_\lambda}\} \right)^2+ \phi \lambda\|\beta\|_1 \right\rbrace \, \text{with}\,\phi\in(0,1]$} & \textcolor{darkpastelgreen}{$\checkmark$}  \rule[-0.8cm]{0pt}{0pt} &  \\
	\hline
	\rule{0pt}{0.5cm} {\LARGE \textcolor{darktangerine}{$\mathbf{\ast}$}} \textbf{Square root LASSO} -- \cite{belloni2011square} & & \multirow{4}{3cm}{\centering \textit{It is not needed to known $\sigma$ to obtain an optimal $\lambda$}} \\
	\rule{0pt}{0.8cm} {\small $\stackbin[\beta]{}{\min} \left\lbrace \left[ \sum_{i=1}^{n}\left(y_i-\sum_{j=1}^{p}x_{ij}\beta_j\right)^2\right]^{1/2} + \lambda\sum_{j=1}^{p}|{\beta_j}| \right\rbrace$} & \textcolor{darkpastelgreen}{$\checkmark$}  \rule[-0.6cm]{0pt}{0pt} & \\
	\hline
	\rule{0pt}{0.5cm} {\LARGE \textcolor{darktangerine}{$\mathbf{\ast}$}} \textbf{Scaled LASSO} -- \cite{sun2012} & & \\
	\rule{0pt}{0.5cm} {\small $\hat{\sigma}\leftarrow \|y-X\hat{\beta}^{old}\|_2/ n^{1/2}, \quad \lambda\leftarrow \hat{\sigma} \lambda_0$} & & \multirow{6}{3cm}{\centering \textit{Simultaneous estimation of $\sigma$ and $\beta$}}\\
	\rule{0pt}{0.8cm} {\small $\hat{\beta}^{new} = \stackbin[\beta]{}{\min} \left\lbrace
	\begin{aligned}
	&x^\top_j(y-X\hat{\beta})/n=\lambda \textrm{sign}(\hat{\beta}_j), \quad &\hat{\beta}_j \not=0,\\
	&x^\top_j(y-X\hat{\beta})/n \in \lambda[-1,1], \quad &\hat{\beta}_j =0.\\
	\end{aligned} 
	\right.$}  & \textcolor{darkpastelgreen}{$\checkmark$}  & \\
	\rule{0pt}{0.6cm} {\small $\hat{\beta}\leftarrow \hat{\beta}^{new}, \; L_{\lambda}(\hat{\beta}^{new}) \leq  L_{\lambda}(\hat{\beta}^{old})$} & & \\
	\rule{0pt}{0.7cm} {\footnotesize $\left(\text{where}\;  L_{\lambda}(\beta)=\frac{\|y-X\beta\|_2^2}{2n}+\lambda \sum_{j=1}^{p} | \beta_j | \right)$}	& \rule[-0.5cm]{0pt}{0pt} & \\
	\hline
	\rule{0pt}{0.5cm}  {\LARGE \textcolor{byzantine}{$\mathbf{\bullet}$}} \textbf{SLOPE} -- \cite{Bogdan2015} & & \multirow{4}{3cm}{\centering \textit{Control of the False Discovery Rate $FDR$}}\\
	\rule{0pt}{0.6cm} {\small $\stackbin[\beta]{}{\min} \left\lbrace \frac{1}{2}\sum_{i=1}^{n}\left(y_i-\sum_{j=1}^{p}x_{ij}\beta_j\right)^2+ \sum_{j=1}^{p} \lambda_j|{\beta}|_{(j)}  \right\rbrace$} & \textcolor{darkpastelgreen}{$\checkmark$}  \rule[-0.7cm]{0pt}{0pt} & \\
	\hline
	\rule{0pt}{0.5cm} {\LARGE \textcolor{darktangerine}{$\mathbf{\ast}$}} \textbf{Debiased LASSO} -- \cite{javanmard2018debiasing} & & \\
	\rule{0pt}{0.6cm} {\small $\hat{\beta}^{debiased}=\hat{\beta}^{L_1}+\frac{1}{n}MX^\top(y-X\hat{\beta}^{L_1}) \sim N(\beta, \sigma^2/n) $} & & \multirow{1}{3cm}{\centering \textit{Characterization of the probability distribution for the estimator of $\beta$ ($\hat{\beta}^{debiased}$)}} \\
	\rule{0pt}{0.4cm} {\footnotesize with  $M=(m_1,\dots, m_p)^\top\in\mathbb{R}^{p\times p}$, where each $m_i\in\mathbb{R}^{p}$ is the solution of} & \multirow{3}*{\textcolor{darkpastelgreen}{$\checkmark$} } &  \\
	\rule{0pt}{0.5cm} {\small $
	\begin{aligned}
	&\min_{m}\; m^\top \hat{\Sigma}m \quad \text{subject to} \; \| \hat{\Sigma}m -e_i \|_{\infty} \leq \mu\\
	\end{aligned} 
	$}  &   & \\
	\rule{0pt}{0.4cm} {\footnotesize ($e_i\in\mathbb{R}^p$ is a standard unit vector, $\hat{\Sigma}=(X^\top X)/n$ and $\mu$ a constraint)}	& \rule[-0.5cm]{0pt}{0pt} & \\
	\hline
	\rule{0pt}{0.5cm} {\LARGE \textcolor{darktangerine}{$\mathbf{\ast}$}} \textbf{LASSO-Zero} -- \cite{descloux2018model} & & \multirow{1}{3cm}{\centering \textit{Excellent trade-off between high true positive rate ($TPR$) and low false discovery rate ($FDR$) }} \\
	\rule{0pt}{0.5cm} {\small $\stackbin[\beta]{}{\min}  \|\beta\|_1 + \|\gamma\|_1$} & \multirow{3}*{\textcolor{red}{\texttimes}}  &  \\
	{\small $\text{subject to} \; y=\tilde{X}\beta + G\gamma$} & & \\
	\rule{0pt}{0.4cm} {\footnotesize ($G\in \mathbb{R}^{n\times q}$ a noise dictionary and $\tilde{X}=(X | G)$)}& \rule[-0.5cm]{0pt}{0pt} & \\
	\hline
	\multicolumn{3}{c}{ } \\
	\caption{Formulated problems to estimate the $\beta$ vector in a linear regression high dimensional framework ($p>n$), showing if the optimization problems are convex (\textcolor{darkpastelgreen}{$\checkmark$}) or not (\textcolor{red}{\texttimes}). Their main advantages in comparison with the LASSO are displayed in column (\textbf{PROS}) and it is showed if they are a weighted version of the LASSO ({\LARGE \textcolor{byzantine}{$\mathbf{\bullet}$}}) or alternatives ({\LARGE \textcolor{darktangerine}{$\mathbf{\ast}$}}). }
	\label{table_other}
\end{longtable}

Next, we display the results of the simulation study comparing the performance of different procedures which have showed suitable properties.

\subsection{Comparison with other approaches by means of a simulation study}\label{comparison}

As we advanced above, we have made a selection of the most relevant methodologies in terms of good qualities as well as reasonable computational time. Because of the nature of the simulation scenarios of Section \ref{simulation_scenarios}, we have discarded some procedures due to their unsuitable characteristics. Moreover, we took into account to choose methodologies with available code in R (\cite{R}), so everyone can make use of them. We have chosen libraries which provide us with enough resources to fit the models, selecting those created for the author's methodology or the most recently updated option in case of doubt. This selection resulted in:

\begin{itemize}
	{\small 
	\item[$-$]{\textbf{LASSO}: \texttt{glmnet}  of \cite{Friedman2010}, last update June 16, 2020.}
	\item[$-$]{\textbf{SCAD}: \texttt{ncvreg} of \cite{ncvreg}, last update July 9, 2020.}
	\item[$-$]{\textbf{Adaptive LASSO (AdapL)}: \texttt{glmnet} of \cite{Friedman2010}, last update June 16, 2020.}
	\item[$-$]{\textbf{Dantzig selector (Dant)}: \texttt{flare} of \cite{flare}, last update August 2, 2019.}
	\item[$-$]{\textbf{Relaxed LASSO (RelaxL)}: \texttt{relaxo} of \cite{relaxo}, last update February 20, 2015.}
	\item[$-$]{\textbf{Square root LASSO (SqrtL)}: \texttt{flare} of \cite{flare}, last update August 2, 2019.}
	\item[$-$]{\textbf{Scaled LASSO (ScalL)}: \texttt{scalreg} of \cite{scalreg}, last update January 25, 2019.}
}
	
\end{itemize}

Now, we analyze the performance of these algorithms in comparison with the LASSO ones (Section \ref{Performance LASSO}), following the scheme introduced in Section \ref{simulation_scenarios}. Furthermore, we compare their results with an innovative procedure which has proved its efficiency, the distance correlation algorithm for variable selection (DC.VS) of \cite{febrero2019variable}. This last makes use of the correlation distance (\cite{Szekely2007}, \cite{szekely2017energy}) to implement an iterative procedure (forward) deciding in each step which covariate enters the regression model. As a consequence, this methodology is tested instead of the usual forward selection because of its selection improvements. A comparison between the forward selection and the LASSO can be found in \cite{hastie2017extended}. For this purpose, the library \texttt{fda.usc} (\cite{fda.usc}) is employed. The complete simulation results are provided in the Supplementary material.

\begin{figure}[htb]\centering
	\includegraphics[width=\linewidth]{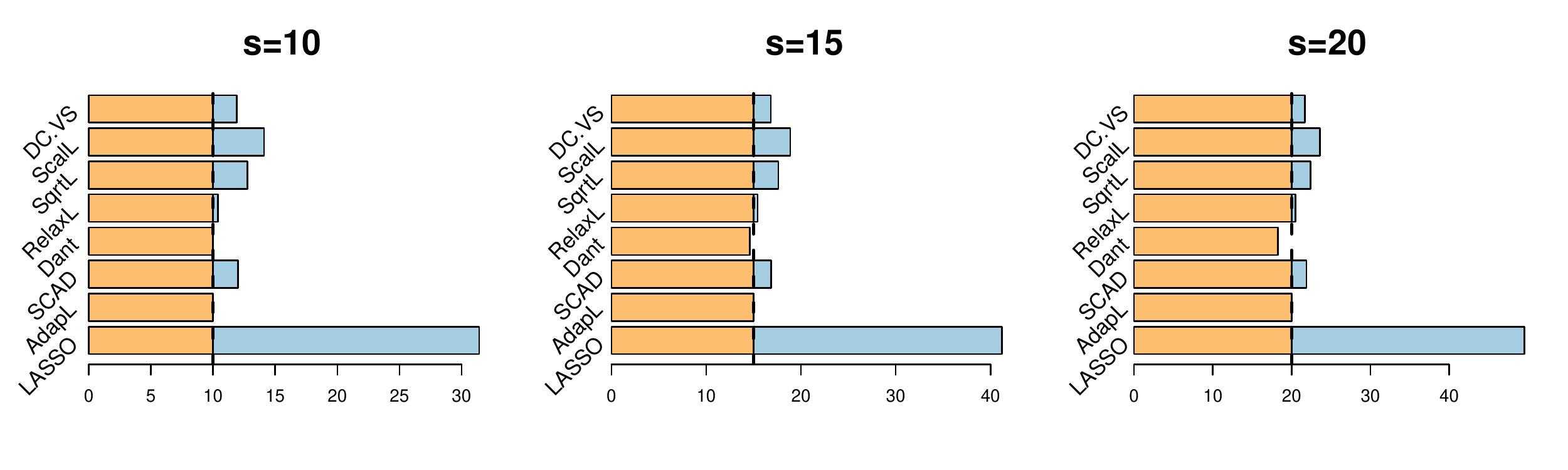}
	\caption{\label{barplots_comparativos_M1} Comparison of the important covariates number (\textcolor{orange}{orange} area) versus noisy ones (\textcolor{bluegray}{blue} area) for $n=400$ in Scenario 1. The dashed line marks the $s$ value.}
\end{figure}

We start studying the easiest framework: the orthogonal design (Scenario 1). In case of simulating under independence between covariates we can see that any of the studied algorithms perform better than the LASSO. These obtain good results searching for the $s$ relevant covariates when $p>n$, and they seem to be able to recover the set $S$ for a large enough value of $n$ (see Figure \ref{barplots_comparativos_M1}). Besides, all of them add less noise to the model and do not overestimate too much the predictions, as the LASSO does. See Table \ref{compare_M1} for a brief comparison. Nevertheless, the only method which selects the complete set $S$ without including any noise to the model, for a large enough value of $n$, is the AdapL algorithm. This last performs incredibly well. The Dant achieves good results in terms of avoiding noise too, however, its convergence to the set $S$ seems more slow. 

\begin{table}[htb] 
	\centering
	\small 
	\begin{tabular}{cccccc}
		\toprule
		\vspace{-0.07cm}
		 \multirow{2}*{\textbf{Scenario}}  & \rule{0pt}{0.35cm} $\boldsymbol{|\hat{S}\cap S|}$ & $\boldsymbol{|\hat{S}\setminus S|}$ & $\boldsymbol{|\hat{S}|}$ &  \textbf{MSE} &  \textbf{\% Dev}\\
		  & {\scriptsize$(15)$} &  &  & {\scriptsize$(2.604)$} & {\scriptsize$(0.9)$} \\  
		\hline
      	\rule{0pt}{0.4cm} \textbf{LASSO} & 15 & 26.2 & 41.2 & 2.091 & 0.919 \\
		\textbf{AdapL} & 15 & 0 & 15 & 2.491 & 0.903 \\ 
		\textbf{SCAD} & 15 & 1.8 & 16.8 & 2.428 & 0.906 \\
		\textbf{Dant} & 14.6 & 0 & 14.6 & 2.957 & 0.885 \\
		\textbf{RelaxL} & 15 & 0.4 & 15.4 & 2.476 & 0.904 \\
		\textbf{SqrtL} & 15 & 2.6 & 17.6 & 2.403 & 0.907 \\ 
		\textbf{ScalL} & 15 & 3.9 & 18.9 & 2.372 & 0.908 \\
		\textbf{DC.VS} & 15 & 1.8 & 16.8 & 2.421 & 0.906 \\ 
		\bottomrule
	\end{tabular}
	\caption{Comparison of all proposed algorithms for Scenario 1 taking $n=400$ and $s=15$. The oracle values are in brackets. }
	\label{compare_M1}
\end{table}

\begin{figure}[htb]\centering
	\includegraphics[width=\linewidth]{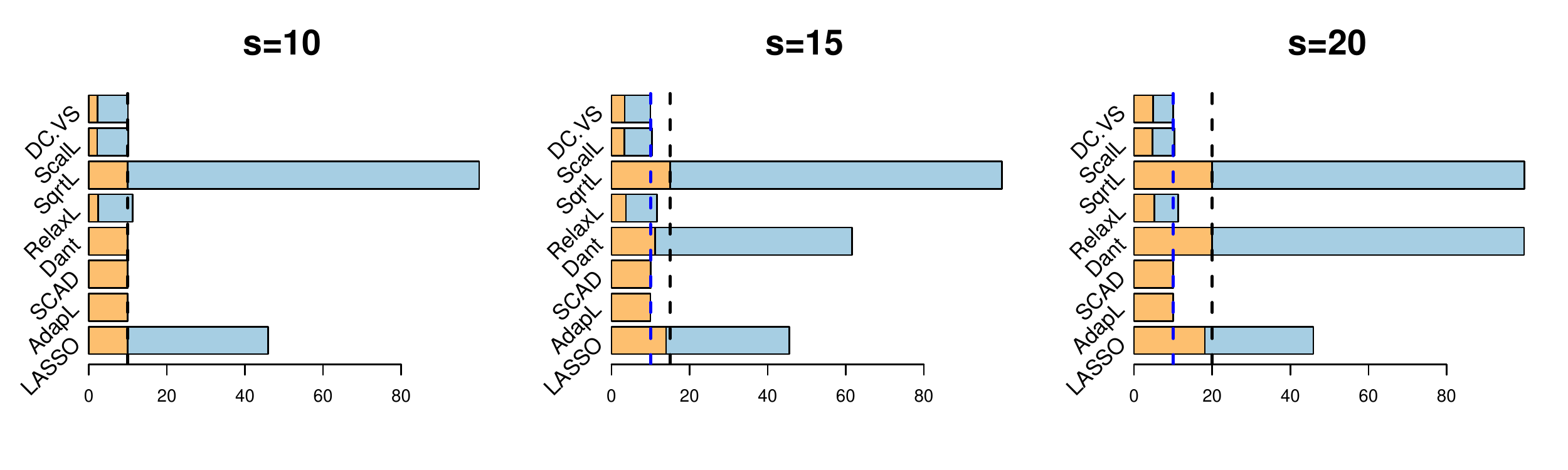}
	\caption{\label{barplots_comparativos_M2_rho_5} Comparison of the important covariates number (\textcolor{orange}{orange} area) versus noisy ones (\textcolor{bluegray}{blue} area) for $n=400$ in Scenario 2 with $\rho=0.5$. The black dashed line marks the considered $s$ value while the blue dashed line where $s=10$.}
\end{figure}

\begin{figure}[htb]\centering
	\includegraphics[width=\linewidth]{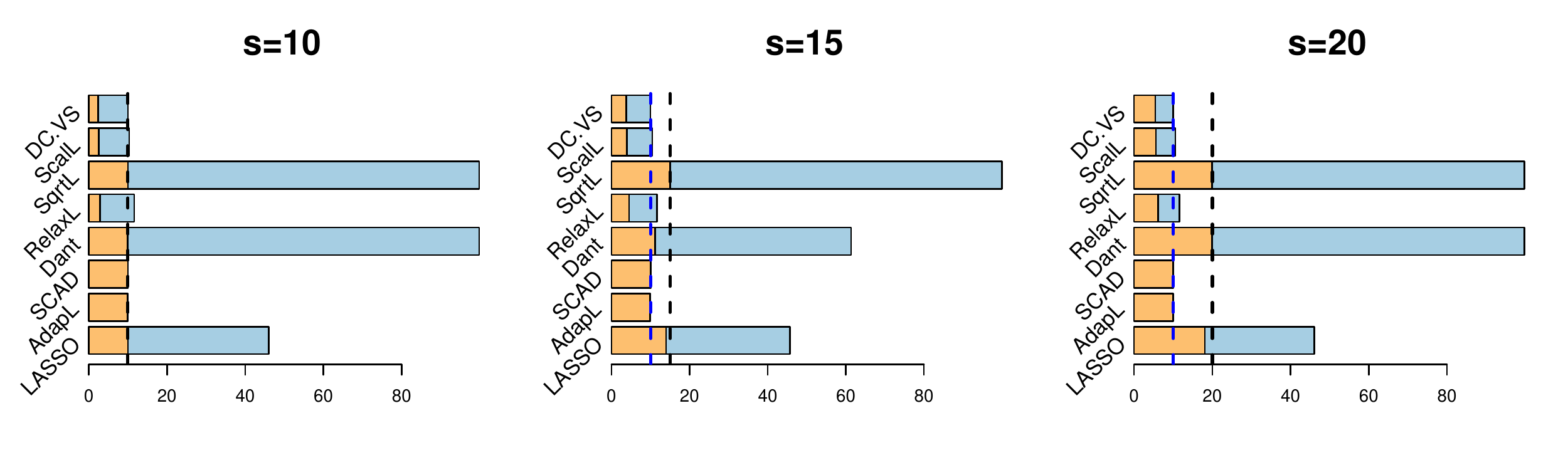}
	\caption{\label{barplots_comparativos_M2_rho_9} Comparison of the important covariates number (\textcolor{orange}{orange} area) versus noisy ones (\textcolor{bluegray}{blue} area) for $n=400$ in Scenario 2 with $\rho=0.9$. The black dashed line marks the considered $s$ value while the blue dashed line where $s=10$.}
\end{figure}

Once we have seen that the proposed alternatives to the LASSO improve the results when there is not a correlation structure between covariates, we want to test their performance under dependence. The first considered model with dependence is the dependence by blocks correlation structure (Scenario 2), simulating a correlation structure of value $\rho$ every ten places. In Section \ref{Performance LASSO} we saw that the LASSO does not select a representative subset of $S$ formed by a bunch of efficient covariates as expected, instead this always tries to recover the complete set adding a lot of noisy ones, which translates in overestimation. A comparative example of all algorithms performance in this scenario, for $s=15$ and $n=400$, is displayed in Table \ref{compare_M2_5} taking $\rho=0.5$ (Scenario 2 with $\rho=0.5$) and in Table \ref{compare_M2_9} simulating with $\rho=0.9$ (Scenario 2 with $\rho=0.9$). Visual examples are showed in Figure \ref{barplots_comparativos_M2_rho_5} and Figure \ref{barplots_comparativos_M2_rho_9} respectively.

\begin{table}[htb] 
	\centering
	\small 
	\begin{tabular}{ccccccc}
		\toprule
		\vspace{-0.07cm}
		\multirow{2}*{$\boldsymbol{\rho}$} & \multirow{2}*{\textbf{Scenario}}  & $\boldsymbol{|\hat{S}\cap S|}$ & \rule{0pt}{0.35cm} $\boldsymbol{|\hat{S}\setminus S|}$ & $\boldsymbol{|\hat{S}|}$ &  \textbf{MSE} &  \textbf{\% Dev}\\
		&  & {\scriptsize$(15)$} &  &  & {\scriptsize$(1.389)$} & {\scriptsize$(0.9)$} \\   
		\hline
		\multirow{8}*{$\boldsymbol{\rho=0.5}$}& \rule{0pt}{0.4cm} \textbf{LASSO} & 14 & 31.6 & 45.6 & 1.346 & 0.949 \\
		& \textbf{AdapL} & 10 & 0 & 10 & 1.346 & 0.949 \\ 
		& \textbf{SCAD} & 10.1 & 0 & 10.1 & 1.346 & 0.949 \\ 
		& \textbf{Dant} & 11.2 & 50.4 & 61.6 & 4.968 & 0.811 \\
		& \textbf{RelaxL} & 3.7 & 8 & 11.7 & 1.377 & 0.947 \\ 
		& \textbf{SqrtL} & 15 & 85 & 100 & 1.346 & 0.949 \\
		& \textbf{ScalL} & 3.4 & 7.1 & 10.4 & 1.374 & 0.948 \\
		& \textbf{DC.VS} & 3.4 & 6.6 & 10 & 1.346 & 0.949  \\
		\bottomrule
	\end{tabular}
	\caption{Comparison of all proposed algorithms for Scenario 2 with $\rho=0.5$ taking $n=400$ and $s=15$. The oracle values are in brackets.}
	\label{compare_M2_5}
\end{table}

\begin{table}[htb] 
	\centering
	\small 
	\begin{tabular}{ccccccc}
		\toprule
		\vspace{-0.07cm}
		\multirow{2}*{$\boldsymbol{\rho}$} & \multirow{2}*{\textbf{Scenario}}  & \rule{0pt}{0.35cm} $\boldsymbol{|\hat{S}\cap S|}$ & $\boldsymbol{|\hat{S}\setminus S|}$ & $\boldsymbol{|\hat{S}|}$ &  \textbf{MSE} &  \textbf{\% Dev}\\
		&  & {\scriptsize$(15)$} &  &  & {\scriptsize$(2.5)$} & {\scriptsize$(0.9)$} \\   
		\hline 
		\multirow{8}*{$\boldsymbol{\rho=0.9}$}& \rule{0pt}{0.4cm} \textbf{LASSO} & 14 & 31.7 & 45.7 & 2.42 & 0.912 \\
		& \textbf{AdapL} & 9.9 & 0.1 & 10 & 2.423 & 0.912 \\         
		& \textbf{SCAD} & 10.1 & 0 & 10.1 & 2.423 & 0.912 \\
		& \textbf{Dant} & 11.1 & 50.2 & 61.3 & 6.013 & 0.781 \\
		& \textbf{RelaxL} & 4.5 & 7.2 & 11.7 & 2.438 & 0.911 \\
		& \textbf{SqrtL} & 15 & 84.9 & 100 & 2.42 & 0.912 \\
		& \textbf{ScalL} & 4 & 6.5 & 10.4 & 2.567 & 0.906 \\
		& \textbf{DC.VS} & 3.8 & 6.2 & 10 & 2.423 & 0.912  \\ 
		\bottomrule
	\end{tabular}
	\caption{Comparison of all proposed algorithms for Scenario 2 with $\rho=0.9$ taking $n=400$ and $s=15$. The oracle values are in brackets.}
	\label{compare_M2_9}
\end{table}

The Dant algorithm as well as the SqrtL suffer from the same issue. We can see as these algorithms are not able of interpreting the structure of the data and select almost the $p$ covariates in some cases. Here, the Dant mimics the performance of the LASSO when there exists the same  correlation between important covariates and noise ones as in the $s=15$ and $s=20$ framework. In these situations, this algorithm recovers $10$ out of the $s$ relevant variables but then, this is unable to distinguish between the rest of important covariates and noise. This seems due to the dependence by blocks structure: important covariates already selected by the model have the same correlation with the rest of relevant ones as with noisy covariates placed every ten locations. Then, the Dant and the SqrtL do not overcome the spurious correlations phenomenon and tend to select too many covariates. Besides, these procedures perform even worse than the LASSO in both frameworks of the Scenario 2 adding more noise and overestimating the prediction accuracy. As a result, we can conclude that both methods are not suitable when we have the same strong correlation between remaining important covariates and noisy ones.

In contrast, the rest of alternatives seem to perform better, trying to select a representative subset of length $10$ approximately. However, not all the remaining procedures select a representative subset between the $s$ important variables. Instead, the majority change relevant covariates for noisy ones quite correlated with the previous ones, covering the complete set $S$. Into words, if a procedure chooses a noise covariate it is expected that this last is a representative of some not included relevant covariate to achieve a good explanation of the data. We can see a proof of this phenomenon for the RelaxL, ScalL and Dist in Section 5.1 of the Supplementary material. Only the AdapL and the SCAD algorithms seem to behave properly in this sense, recovering $10$ elements of the set $S$. All these methodologies correct a bit the overestimation produced by the LASSO.

\begin{figure}[htb]\centering
	\includegraphics[width=\linewidth]{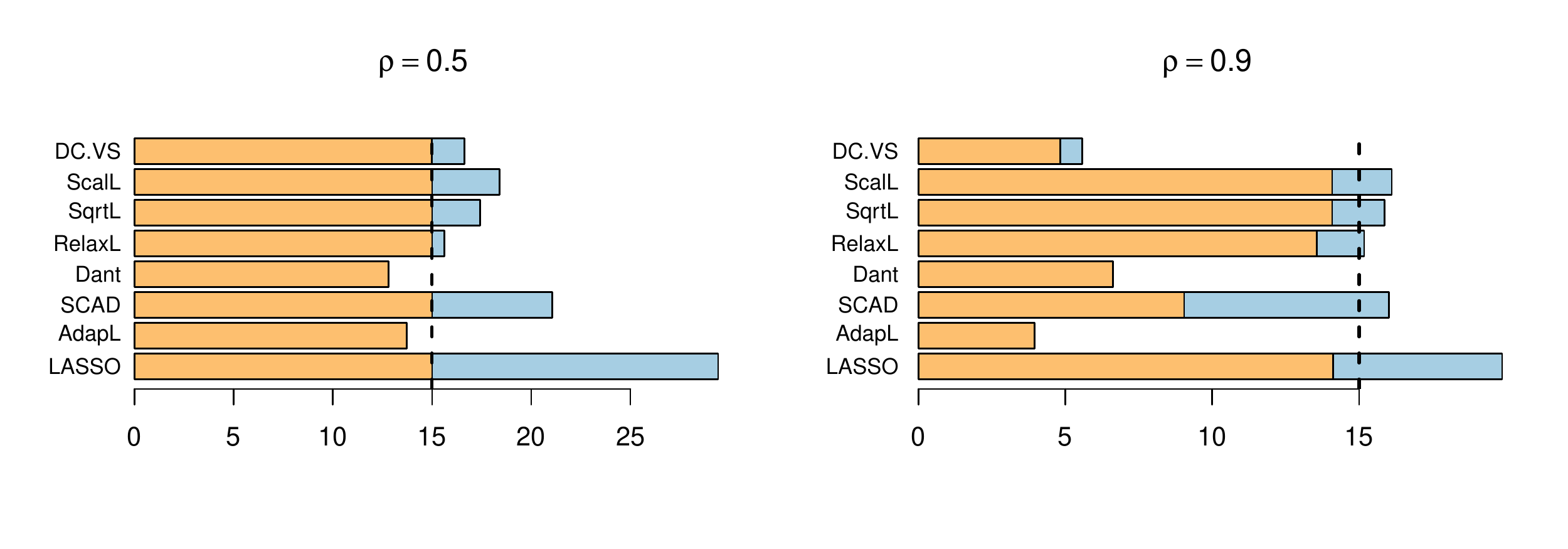}
	\caption{\label{barplots_comparativos_M3_s1} Comparison of the important covariates number (\textcolor{orange}{orange} area) versus noisy ones (\textcolor{bluegray}{blue} area) for $n=400$ in Scenario 3.a. The dashed line marks the $s$ value.}
\end{figure}

Finally, we simulate under the Toeplitz covariance structure of Scenario 3. We consider a first scenario, where the relevant covariates are located in the first $s=15$ placements (Scenario 3.a), and a second one, where we simulate only $s=10$ important variables and they are placed every ten sites (Scenario 3.b). Hence, we expect for the Scenario 3.a to obtain a representative subset of the set $S$, with cardinal less than $s$ as we explained in Section \ref{Performance LASSO}. Specially, when the correlation between covariates is strong, as for $\rho=0.9$. It is owing to the fact that we have, in this scenario, several relevant covariates with a representative correlation between them. Roughly speaking, because of the Toeplitz covariance structure, one variable could be ``easily'' explained by others in its neighborhood. This translates in the possibility of interchanging last variables of $S$ with nearby ones. Then, for $\rho=0.5$, because $0.5^5 \leq 0.05$, we consider as good representatives those covariates which distance is less than $4$ to some position of $S$. When $\rho=0.9$ this distance is enlarged and there are many more possibilities. In contrast, simulating the Scenario 3.b, we would expect the algorithm to select all the $10$ relevant covariates in the best case or a representative subset following this criteria.

\begin{table}[htb] 
	\centering
	\small 
	\begin{tabular}{ccccccc}
		\toprule
		\vspace{-0.07cm}
		\multirow{2}*{$\boldsymbol{\rho}$} & \multirow{2}*{\textbf{Scenario}}  & $\boldsymbol{|\hat{S}\cap S|}$ & \rule{0pt}{0.35cm} $\boldsymbol{|\hat{S}\setminus S|}$ & $\boldsymbol{|\hat{S}|}$ &  \textbf{MSE} &  \textbf{\% Dev}\\
		&  & {\scriptsize$(15)$} &  &  & {\scriptsize$(1.139)$} & {\scriptsize$(0.9)$} \\  
		\hline
		\multirow{8}*{$\boldsymbol{\rho=0.5}$}& \rule{0pt}{0.4cm} \textbf{LASSO} & 15 & 14.4 & 29.4 & 0.972 & 0.914 \\
		& \textbf{AdapL} & 13.7 & 0 & 13.7 & 1.195 & 0.894 \\
		& \textbf{SCAD} & 15 & 6.1 & 21.1 & 1.016 & 0.910 \\
		& \textbf{Dant} & 12.8 & 0 & 12.8 & 1.443 & 0.873 \\
		& \textbf{RelaxL} & 15 & 0.6 & 15.6 & 1.078 & 0.905 \\
		& \textbf{SqrtL} & 15 & 2.4 & 17.4 & 1.053 & 0.907 \\
		& \textbf{ScalL} & 15 & 3.4 & 18.4 & 1.039 & 0.908 \\
		& \textbf{DC.VS} & 15 & 1.6 & 16.6 & 1.061 & 0.906 \\
		\bottomrule
	\end{tabular}
	\caption{Comparison of all proposed algorithms for Scenario 3.a taking $n=400$ and $\rho=0.5$. The oracle values are in brackets.}
	\label{compare_M3_5}
\end{table}

For the Scenario 3.a we can appreciate in Figure \ref{barplots_comparativos_M3_s1} a similar phenomenon as the one observed in Scenario 2. This is translated in the existence of algorithms which try to recover the complete set $S$, like the LASSO, the SCAD, the RelaxL, the SqrtL or the ScalL. But, the rest of algorithms, the AdapL, the Dant algorithm and the DC.VS, search for a representative subset without including noise. A summary of their performance for $\rho=0.5$ is displayed In Table \ref{compare_M3_5} and for $\rho=0.9$ In Table \ref{compare_M3_9}. Taking $\rho=0.5$ we appreciate that the AdapL and the Dant are the only procedures which select the number of efficient covariates needed to explain, at least, the $90\%$ of the covariance. A similar behavior could be considered for the DC.VS but this adds more noise and selects more than $s=15$ covariates for $\rho=0.5$. In Section 5.2.1 of the Supplementary material the percentage of times the relevant covariates are selected for these algorithms is displayed.

\begin{table}[htb] 
	\centering
	\small 
	\begin{tabular}{ccccccc}
		\toprule
		\vspace{-0.07cm}
		\multirow{2}*{$\boldsymbol{\rho}$} & \multirow{2}*{\textbf{Scenario}}  & \rule{0pt}{0.35cm} $\boldsymbol{|\hat{S}\cap S|}$ & $\boldsymbol{|\hat{S}\setminus S|}$ & $\boldsymbol{|\hat{S}|}$ &  \textbf{MSE} &  \textbf{\% Dev}\\
		&  & {\scriptsize$(15)$} &  &  & {\scriptsize$(3.80723)$} & {\scriptsize$(0.9)$} \\   
		\hline 
		\multirow{8}*{$\boldsymbol{\rho=0.9}$}& \rule{0pt}{0.4cm} \textbf{LASSO} & 14.1 & 5.8 & 19.9 & 3.620 & 0.908 \\
		& \textbf{AdapL} & 4 & 0 & 4 & 4.543 & 0.885\\                  
		& \textbf{SCAD} & 9.1 & 7 & 16 & 3.658 & 0.907 \\
		& \textbf{Dant} & 6.6 & 0 & 6.6 & 4.92 & 0.875 \\
		& \textbf{RelaxL} & 13.6 & 1.6 & 15.2 & 3.728 & 0.906 \\
		& \textbf{SqrtL} & 14.1 & 1.8 & 15.9 & 3.717 & 0.906 \\
		& \textbf{ScalL} & 14.1 & 2 & 16.1 & 3.701 & 0.906 \\
		& \textbf{DC.VS} & 4.8 & 0.8 & 5.6 & 4.285 & 0.891 \\ 
		\bottomrule
	\end{tabular}
	\caption{Comparison of all proposed algorithms for Scenario 3.a taking $n=400$ and $\rho=0.9$. The oracle values are in brackets.}
	\label{compare_M3_9}
\end{table}

\begin{figure}[htb]\centering
	\includegraphics[width=\linewidth]{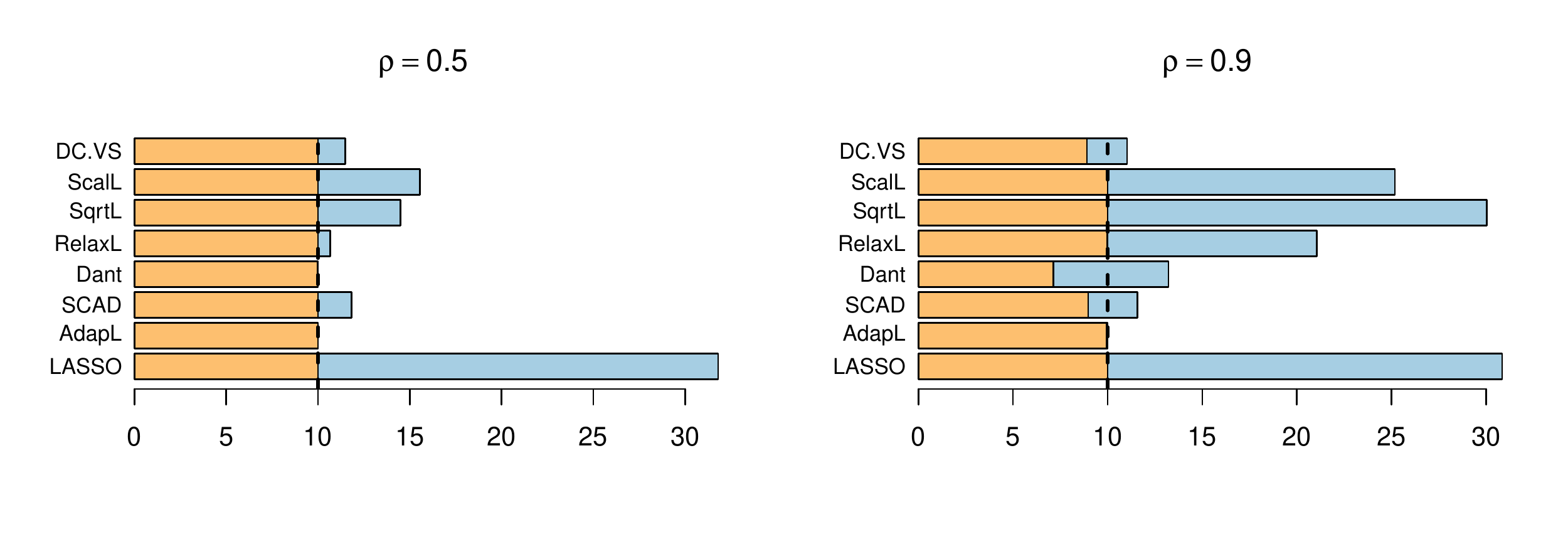}
	\caption{\label{barplots_comparativos_M3_s2} Comparison of the important covariates number (\textcolor{orange}{orange} area) versus noisy ones (\textcolor{bluegray}{blue} area) for $n=400$ in Scenario 3.b. The dashed line marks the $s$ value.}
\end{figure}

 Studying the provided results for $\rho=0.9$ (Table \ref{compare_M3_9}) we can claim that Dist achieves the best results in terms of prediction when the correlation is large, but this pays the price of including more irrelevant information than the adaptive LASSO or the Dant. In contrast, when $\rho=0.5$, the selection of covariates made by the DC.VS results in an overestimation of the model. Finally, if we compare the AdapL with the Dant algorithm results, we see that it seems like the first one obtains a better trade-off between selection of covariates and estimation. We notice that, for $\rho=0.9$, the AdapL selects less covariates of $S$ but achieves a better performance in terms of explanation of the data. All these three approaches select less than $s=15$ covariates but a number large enough to guarantee a good explanation of the covariance.

At this point, it is interesting to notice that the Dant performs correctly in this dependence context in comparison with the Scenario 2 framework. Now, this is able to recover a representative subset of $S$ without adding noise to the model. This phenomenon could be explained taking into account that in Scenario 3.a we have not too many noisy covariates highly correlated with the ones of $S$, specially for $\rho=0.5$. Only those in the neighborhood of the $15^{th}$ could be a threat. However, in the Scenario 2 we had covariates correlated ten by ten and, as a result, every relevant covariate is correlated with 8 irrelevant ones at least. In contrast, the SqrtL keeps its bad behavior and the SCAD algorithm starts to perform poorly. This last brings out the fact that the SCAD procedure suffers when all the covariates are correlated among them. This happens when the important covariates are close in location as in the case of Scenario 3.a, however, when these are more scattered, like in Scenario 3.b, the algorithm performs better.

Eventually, we compare the results obtained for the Scenario 3.b. An example is displayed in Figure \ref{barplots_comparativos_M3_s2} and the rest of results are provided in the Supplementary material. Simulating for $\rho=0.5$ we observe that all the proposed algorithms outperform the LASSO results. These procedures try to recover the complete set $S$ as it is expected taking into account that the number of efficient covariates is $10$ now. Nevertheless, when $\rho=0.9$, some drawbacks come up. Some of them interchange relevant covariates with irrelevant ones quite correlated with these. This is product of the strong correlation structure of the Toeplitz covariance. These are the SCAD and the DC.VS algorithms. Maybe, we can include in this last group the Dant, although it is doubtful. Section 5.2.2 of the Supplementary material collects the percentage of times a representative of the $10$ relevant covariates enters the model. Other procedures, like the RelaxL, the SqrtL or the ScalL add unnecessary noise, overestimating the model. Only one algorithm is almost capable of recovering the $s$ variables without adding more noise to the model, this is the AdapL algorithm. All the alternatives correct the overestimation in the prediction made by the LASSO though.


\section{Discussion: some guidance about LASSO}\label{conclusions}

Currently, the LASSO regression keeps being a broadly employed covariates selection technique. Despite its several advantages, some strict necessary requirements could make difficult a correct performance of this methodology, as it was explained in Section \ref{sec:LASSO}. As we argued at the beginning of the document, there are no global recommendations about the use of the LASSO in terms of the nature of the data or when some of these conditions do not hold. With the aim of shed light on this topic, we have analyzed the LASSO drawbacks, studying modifications and alternatives to overcome these. Besides, an extensive simulation study has been carried out to illustrate the behavior of LASSO in the best possible scenario and in trickier ones carefully chosen (Section \ref{simulation_scenarios}), comparing this with the one of recent modifications and other alternatives (Section \ref{other}). In view of the results, we give next some guidance on how to choose a proper covariates selector according to the nature of data.

We have seen that even in scenarios where there no exists dependence, the LASSO procedure performs poorly in the sense that adds more noise than relevant covariates to the model. Nevertheless, this recovers the complete set $S$ paying the price of noise addition. As a result, this selection of covariates overestimates the prediction errors. These drawbacks can be easily overcome making use of other penalization techniques, keeping the ideas of the $L_1$ regularization, as the ones proposed in Section \ref{comparison}. All these procedures improve the LASSO results in this independence context, decreasing the number of selected noisy covariates and correcting the overestimation. We highlight the adaptive LASSO (\cite{Zou2006}), the relaxed LASSO (\cite{meinshausen2007relaxed}), the Dantzig selector (\cite{candes2007dantzig}) and the distance correlation algorithm (\cite{febrero2019variable}) as the best of the proposed algorithms for this framework. They are able to recover the complete set $S$ adding little noise for a great enough value of $n$. Besides, they correct the prediction errors.

These disadvantages of the LASSO are also transferred to dependence structures. The confusion phenomenon appears in these situations involving an increment of false discoveries and overestimation. Here, not all the proposed methods of Section \ref{comparison} performs properly. It depends on the nature of the correlation which methodologies will be efficient. In order to test their adequacy, we have consider different scenarios under a dependence by blocks structure and under a time series one. We found that the adaptive LASSO (\cite{Zou2006}) and the distance correlation algorithm (\cite{febrero2019variable}) are the only ones quite competent in all these scenarios, regarding to different types of dependence.

 The quality of some procedures performance vary according to the type of correlation structure of the data. Examples of this are the SCAD penalization (\cite{fan1997comments}) and the Dantzig selector (\cite{candes2007dantzig}). The first one achieves a good performance except for the case when there exists strong correlations between all the relevant covariates. In contrast, the Dantzig selector performs properly in these scenarios, but this is not capable of recovering the important covariates, avoiding noise, under a dependence structure by blocks.
 
  The rest of analyzed methods: relaxed LASSO, square root LASSO and scaled LASSO, present a deficient behavior when there exists some type of dependence structure between the covariates. In case of the dependence by blocks, as in Scenario 2, the relaxed LASSO and the scaled LASSO mix relevant covariates with unimportant ones even for $\rho=0.5$, whereas the square root LASSO does not take advantage of the correlation structure. For the Toeplitz covariance scenario, all of them mimic the LASSO behavior trying to recover the complete set $S$ instead of making use of the structure of the data to correctly adjust the regression model.

  As mentioned in Section \ref{simulation_scenarios}, in the different considered dependence structures, all covariates are in the same scale. Analysis about the effect of different scales on the covariates, combined with dependence structures, are interesting for future work.

\section*{Acknowledgment}

This work has been partially supported by the Spanish Ministerio
de Econom\'ia, Industria y Competitividad grant MTM2016-76969-P, Xunta de Galicia Competitive Reference Groups 2017-2020 (ED431C 2017/38) and the Xunta de Galicia grant ED481A-2018/264.


\bibliographystyle{apalike}
\bibliography{Libreria_LASSO}

\pagebreak

\appendix
\section{Figures}

\begin{figure}[htb]\centering
	\includegraphics[width=\linewidth]{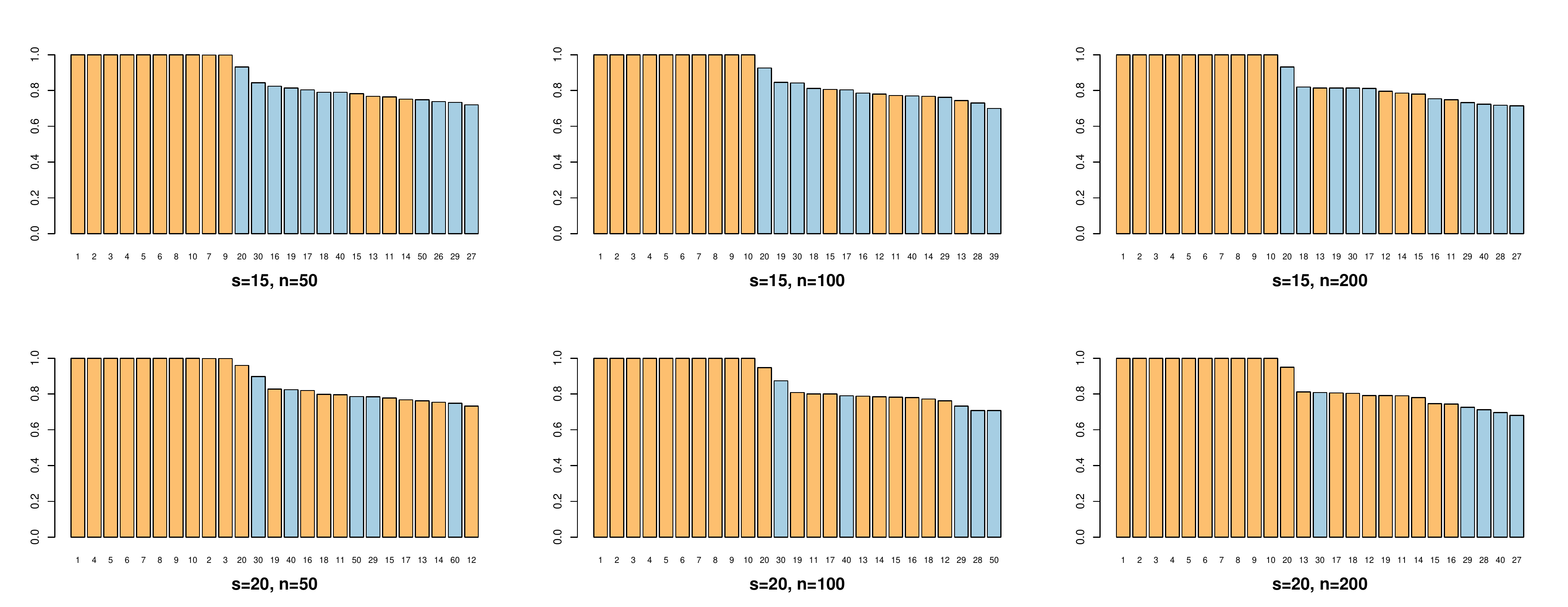}
	\caption{\label{barplots_M2_25_rho_5} The first 25 covariates with highest selection probability for the LASSO in Scenario 2 with $\rho=0.5$. The important covariates of the model are in \textcolor{orange}{orange} while the noisy ones in \textcolor{bluegray}{blue}.}
\end{figure}

\begin{figure}[htb]\centering
	\includegraphics[width=\linewidth]{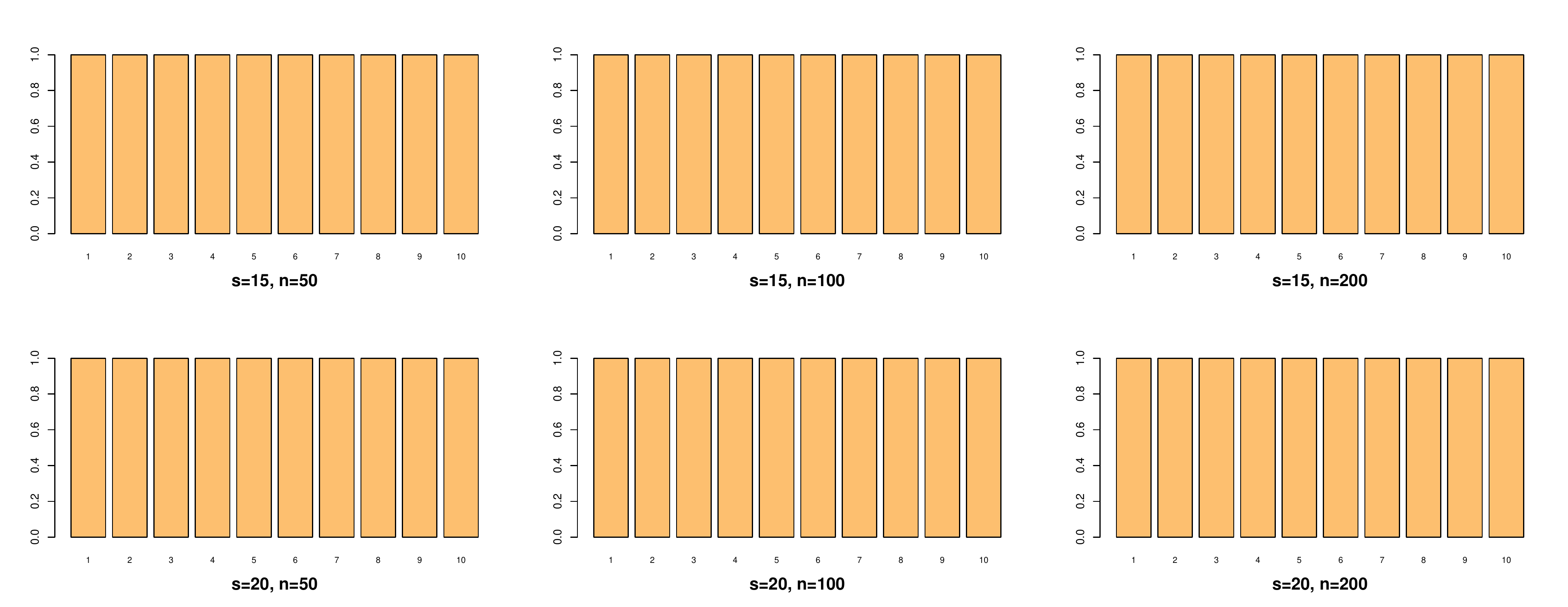}
	\caption{\label{barplots_M2_clases_rho_5} Percentage of times a representative of the $10$ first covariates enters the model in Scenario 2 with $\rho=0.5$ for the LASSO.}
\end{figure}

\begin{figure}[htb]\centering
	\includegraphics[width=\linewidth]{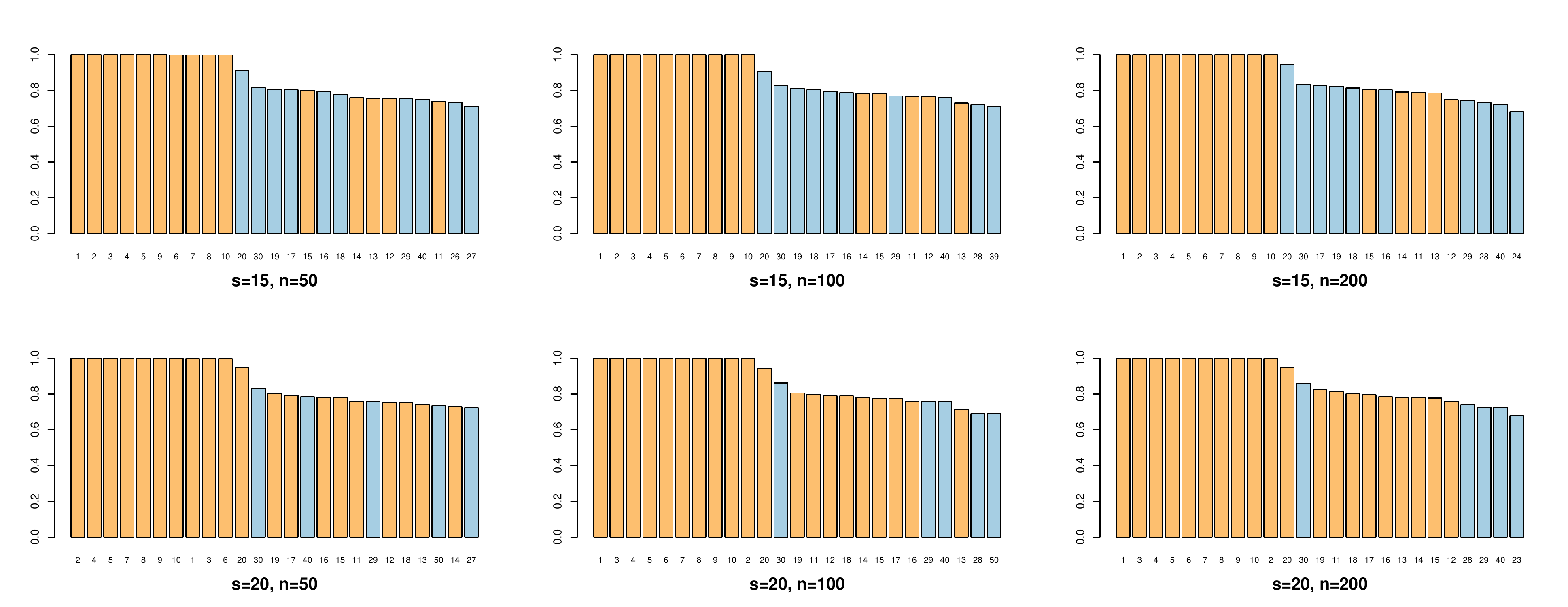}
	\caption{\label{barplots_M2_25_rho_9} The first 25 covariates with highest selection probability for the LASSO in Scenario 2 with $\rho=0.9$. The important covariates of the model are in \textcolor{orange}{orange} while the noisy ones in \textcolor{bluegray}{blue}.}
\end{figure}

\begin{figure}[htb]\centering
	\includegraphics[width=\linewidth]{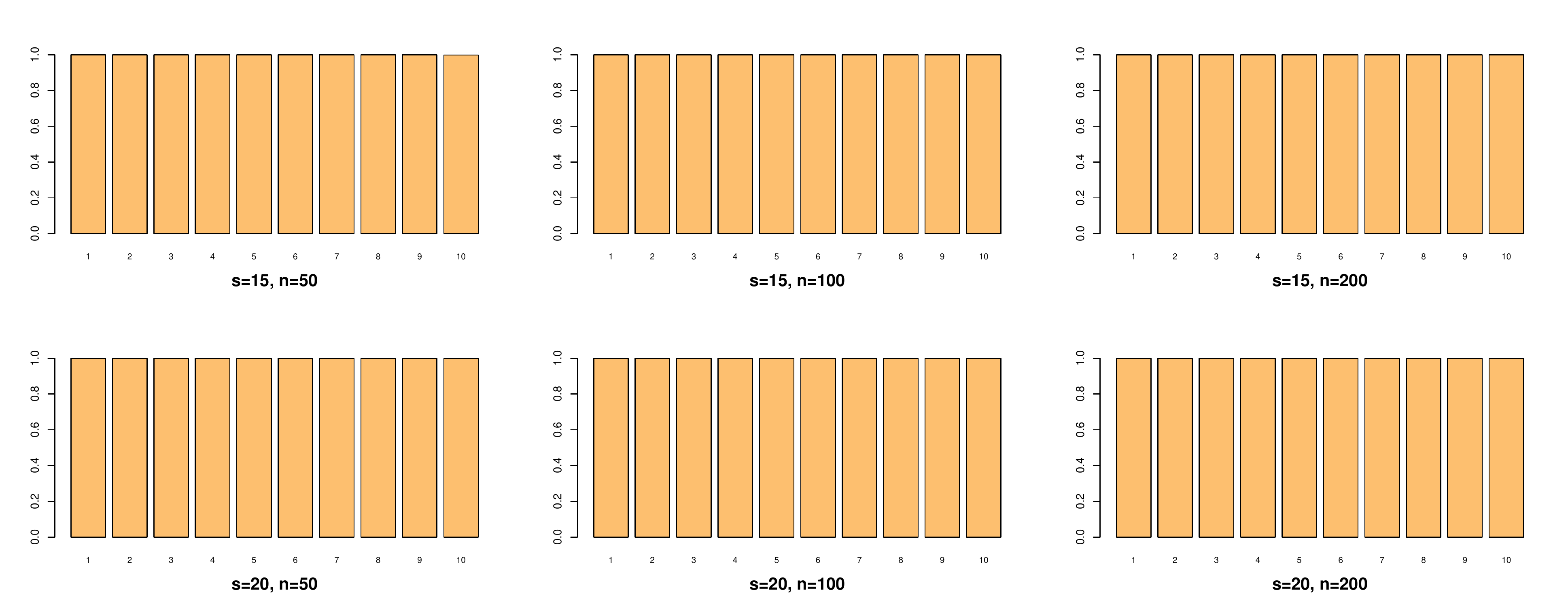}
		\caption{\label{barplots_M2_clases_rho_9} Percentage of times a representative of the $10$ first covariates enters the model in Scenario 2 with $\rho=0.9$ for the LASSO.}
\end{figure}

\end{document}